\documentclass[a4paper,reqno,12pt]{amsart}
\usepackage[all]{xy}           %For commutative diagrams
\usepackage{amssymb}           %For double arrows
\usepackage[utf8]{inputenc}
\usepackage[T1]{fontenc}
\usepackage{hyperref}
\usepackage{tikz-cd}
\usepackage{eucal}
\usepackage{epsfig}
\usepackage{pst-grad} % For gradients
\usepackage{pst-plot} % For axes
\numberwithin{equation}{section}

\newtheorem{definition}{Definition}[section]

\newtheorem{theorem}[definition]{Theorem}
\newtheorem{proposition}[definition]{Proposition}
\newtheorem{corollary}[definition]{Corollary}
\newtheorem{remarkth}[definition]{Remark}
\newtheorem{example}[definition]{Example}
\newenvironment{remark}{\begin{remarkth}\upshape}{\hfill$\diamond$\end{remarkth}}

\renewcommand{\emph}[1]{{\bfseries\itshape{#1}}}

\makeatletter
\newcommand\prol{\@ifstar{\@proldf}{\@prolpf}}  %% if * dual else primal
\def\@prolpf{\@ifnextchar[{\@prolpf@wrt}{\@prolpf@}}
\def\@prolpf@wrt[#1]#2{\@ifnextchar[{\@prolpf@wrt@at{#1}{#2}}{\@prolpf@wrt@{#1}{#2}}}
\def\@prolpf@wrt@at#1#2[#3]{\prolsymbol^{#1}_{#3}#2}
\def\@prolpf@wrt@#1#2{\prolsymbol^{#1}#2}
\def\@prolpf@#1{\@ifnextchar[{\@prolpf@at{#1}}{\@prolpf@@{#1}}}
\def\@prolpf@at#1[#2]{\prolsymbol_{#2}#1}
\def\@prolpf@@#1{\prolsymbol#1}
\def\@proldf{\@ifnextchar[{\@proldf@wrt}{\@proldf@}}
\def\@proldf@wrt[#1]#2{\@ifnextchar[{\@proldf@wrt@at{#1}{#2}}{\@proldf@wrt@{#1}{#2}}}
\def\@proldf@wrt@at#1#2[#3]{\prolsymbol^{*#1}_{#3}#2}
\def\@proldf@wrt@#1#2{\prolsymbol^{*#1}#2}
\def\@proldf@#1{\@ifnextchar[{\@proldf@at{#1}}{\@proldf@@{#1}}}
\def\@proldf@at#1[#2]{\prolsymbol^*_{#2}#1}
\def\@proldf@@#1{\prolsymbol^*#1}
\def\prolsymbol{\mathcal{T}}
\makeatother

\begin{document}

\title[On the homogeneity of non-uniform material bodies]{On the homogeneity of non-uniform material bodies}

\author[V. M. Jim\'enez]{V\'ictor Manuel Jim\'enez}
\address{V\'ictor Manuel Jim\'enez:
Instituto de Ciencias Matem\'aticas (CSIC-UAM-UC3M-UCM),
c$\backslash$ Nicol\'as Cabrera, 13-15, Campus Cantoblanco, UAM
28049 Madrid, Spain} \email{victor.jimenez@icmat.es}

\author[M. de Le\'on]{Manuel de Le\'on}
\address{Manuel de Le\'on: Instituto de Ciencias Matem\'aticas (CSIC-UAM-UC3M-UCM),
c$\backslash$ Nicol\'as Cabrera, 13-15, Campus Cantoblanco, UAM
28049 Madrid, Spain} \email{mdeleon@icmat.es}

\author[M. Epstein]{Marcelo Epstein}
\address{Marcelo Epstein:
Department of Mechanical Engineering. University of Calgary. 2500 University Drive NW, Calgary, Alberta, Canada, T2N IN4} \email{epstein@enme.ucalgary.ca}

\keywords{smooth distribution, singular foliation, groupoid, uniformity, homogeneity, material groupoid, material distribution}
\thanks{This work has been partially supported by MINECO Grants  MTM2016-76-072-P and the ICMAT Severo Ochoa projects SEV-2011-0087 and SEV-2015-0554. V.M. Jim\'enez wishes to thank MINECO for a FPI-PhD Position.}
 \subjclass[2000]{}

\begin{abstract}
A groupoid $\Omega \left( \mathcal{B} \right)$ called material groupoid is naturally associated to any simple body $\mathcal{B}$ (see \cite{MAREMDL,FGM2,MEPMDLSEG}). The material distribution is introduced due to the (possible) lack of differentiability of the material groupoid (see \cite{MD,CHARDIST}). Thus, the inclusion of these new objects in the theory of material bodies opens the possibility of studying non-uniform bodies. As an example, the material distribution and its associated singular foliation result in a rigorous and unique subdivision of the material body into strictly smoothly uniform sub-bodies, laminates, filaments and isolated points. Furthermore, the material distribution permits us to present a ``measure" of uniformity of a simple body as well as more general definitions of homogeneity for non-uniform bodies.
\end{abstract}

\maketitle

\tableofcontents

\section{Introduction}
\label{sec:1}
As it is well-known, there exists a close relation between Continuum Mechanics and Differential Geometry since a continuum is described as a differential manifold. Walter Noll \cite{WNOLL} showed that we can obtain an additional differential geometric structure from the mechanical response associated to the body. Noll's work was extended by Wang \cite{CCWANSEG} and Bloom \cite{FBLOO}. Later, a formalism on $G-$structures \cite{MELZA,FBLOO} was presented by M. El\.zanowski and others as a natural framework for uniform bodies. Furthermore, for general (not necessarily uniform) bodies, groupoids and smooth distributions \cite{MEPMDLSEG, MD, CHARDIST} was shown to be useful tools to express in geometrical terms the mechanical properties derived from the constitutive law.\\
\indent{By simplicity, in this paper we will consider a \textit{simple material} $\mathcal{B}$, i.e., the mechanical response $W$ at each point depends on the deformation gradient alone (and not on higher gradients). So, the \textit{material groupoid} $\Omega \left( \mathcal{B} \right)$ over $\mathcal{B}$ consists of all linear isomorphisms $P$ between the tangent spaces $T_{X} \mathcal{B}$ and $T_{Y} \mathcal{B}$ such that}
$$ W \left( F  P , X \right) = W \left( F, Y \right),$$
for any deformation gradient $F$ at $Y$, where $X,Y$ run along the body $\mathcal{B}$. Thus, we realize that $\mathcal{B}$ is (smoothly) uniform if, and only if, $\Omega \left( \mathcal{B} \right)$ is a transitive (Lie) subgroupoid of $\Pi^{1} \left( \mathcal{B} , \mathcal{B} \right)$, where $\Pi^{1} \left( \mathcal{B} , \mathcal{B} \right)$ is the Lie groupoid over $\mathcal{B}$, called \textit{$1-$jets groupoid on $\mathcal{B}$}, of all linear isomorphisms $P$ between the tangent spaces $T_{X} \mathcal{B}$ and $T_{Y} \mathcal{B}$, for $X,Y \in \mathcal{B}$.\\
\indent{In general, $\Omega \left( \mathcal{B} \right)$ is not a Lie subgroupoid of $\Pi^{1} \left( \mathcal{B} , \mathcal{B} \right)$. To deal with this problem we have introduced the \textit{material distribution} $A \Omega^{T} \left( \mathcal{B} \right)$ (see \cite{MD} or \cite{CHARDIST}). $A \Omega^{T} \left( \mathcal{B} \right)$ is generated by the (local) left-invariant vector fields on $\Pi^{1} \left( \mathcal{B} ,\mathcal{B} \right)$ which are in the kernel of $TW$. Due to the groupoid structure, we can still associate two new objects to $A \Omega^{T} \left( \mathcal{B} \right)$, denoted by $A \Omega \left( \mathcal{B} \right)$ and $A \Omega^{\sharp} \left( \mathcal{B} \right)$, as defined by the following diagram:}

\begin{large}
\begin{center}
 \begin{tikzcd}[column sep=huge,row sep=huge]
\Pi^{1} \left( \mathcal{B} , \mathcal{B} \right)\arrow[r, "A \Omega^{T} \left( \mathcal{B} \right)"] &\mathcal{P} \left( T \Pi^{1} \left( \mathcal{B} , \mathcal{B} \right) \right) \arrow[d, "T\alpha"] \\
 \mathcal{B} \arrow[u,"\epsilon"] \arrow[r,"A \Omega^{\sharp} \left( \mathcal{B} \right)"] \arrow[ru,dashrightarrow, "A \Omega \left( \mathcal{B} \right)"]&\mathcal{P} \left( T \mathcal{B} \right)
 \end{tikzcd}
\end{center}
\end{large}
\noindent{Here $\mathcal{P} \left( E \right)$ defines the power set of $E$, $\epsilon \left( X \right)$ is the identity map of $T_{X}  \mathcal{B}$ and $\alpha : \Pi^{1} \left( \mathcal{B} , \mathcal{B} \right) \rightrightarrows \mathcal{B}$ denotes the source map.}\\
\indent{From its definition, the distributions $A  \Omega^{T} \left( \mathcal{B} \right)$ and $A \Omega^{\sharp} \left( \mathcal{B} \right)$, are integrable (in the sense of Stefan \cite{PS} and Sussmann \cite{HJS}), and they provide two foliations, $\overline{\mathcal{F}}$ on $\Pi^{1} \left( \mathcal{B}, \mathcal{B} \right)$ and $\mathcal{F}$ on $\mathcal{B}$, such that $\Omega \left( \mathcal{B}  \right)$ is union of leaves of $\overline{\mathcal{F}}$. As a consequence, we have that $\mathcal{B}$ can be covered by a foliation of some kind of smoothly uniform ``sub-bodies'' (these submanifolds are not sub-bodies in the usual sense of continuum mechanics \cite{CCWAN} because the dimensions are not, necessarily, equal to the dimension of $\mathcal{B}$), called \textit{material submanifolds}. The material distribution also offers a tool apt to provide a general classification of smoothly non-uniform bodies and opens the possibility to distinguish various degrees of uniformity. In addition, homogeneity may be generalized in such a way that any simple body can be tested to be homogeneous. A first step in this direction was done in \cite{MGEOEPS} where the authors give a homogeneity condition for bundle and laminated bodies.}\\

\indent{The paper is structured as follows: Section 1 is devoted to give a very brief introduction to groupoids. Here we present, as an example, the groupoid of $1-$jets of local automorphisms on a manifold $M$. In the next section we study the characteristic distribution, a general smooth distribution associated to any subgroupoid of a Lie groupoid. Then, in Section 3 we apply this construction to the theory of simple bodies generating in this way the so called material groupoid and material distribution. By using these two mathematical objects we remind the concept of \textit{graded uniformity} of a simple body. Section 4 is used to introduced a new definition of homogeneity for non-uniform bodies which generalize the known definition for smoothly uniform bodies. Some characterizations are given related with the integrability of the material groupoid and the material distribution. Finally, we study two examples of non-uniform body in which the homogeneity is checked.}
\section{Groupoids}
\label{sec:2}
First, we shall give a brief introduction to \textit{Lie groupoids}. The standard reference on groupoids is \cite{KMG}; for a short introduction there is book \cite{JNM} (written in Spanish) which can be useful.  Other recommendable references are \cite{EPSBOOK} and \cite{WEINSGROUP}.

\begin{definition}
\rm
Let $ M$ be a set. A \textit{groupoid} over $M$ is given by a set $\Gamma$ provided with the following maps: $\alpha,\beta : \Gamma \rightarrow M$ (\textit{source} and \textit{target maps}, respectively), $\epsilon: M \rightarrow \Gamma$ (\textit{identities map}), $i: \Gamma \rightarrow \Gamma$ (\textit{inversion map}) and   
$\cdot : \Gamma_{\left(2\right)} \rightarrow \Gamma$ (\textit{composition law}) where for each $k \in \mathbb{N}$,
$$\Gamma_{\left(k\right)} := \{ \left(g_{1}, \hdots , g_{k}\right) \in \Gamma^{k}  \ : \ \alpha\left(g_{i}\right)=\beta\left(g_{i+1}\right), \ i=1, \hdots , k -1 \},$$ satisfying the following properties:\\
\begin{itemize}
\item[(1)] $\ \ $ $\alpha$ and $\beta$ are surjective and, for each $\left(g,h\right) \in \Gamma_{\left(2\right)}$, we have
$$ \alpha\left(g \cdot h \right)= \alpha\left(h\right), \ \ \ \beta\left(g \cdot h \right) = \beta\left(g\right).$$
\item[(2)] $\ \ $ Associativity of the composition law, i.e.,
$$ g \cdot \left(h \cdot k\right) = \left(g \cdot h \right) \cdot k, \ \forall \left(g,h,k\right) \in \Gamma_{\left(3\right)}.$$
\item[(3)] $\ \ $ For all $ g \in \Gamma$,
$$ g \cdot \epsilon \left( \alpha\left(g\right)\right) = g = \epsilon \left(\beta \left(g\right)\right)\cdot g .$$
In particular,
$$ \alpha \circ  \epsilon \circ \alpha = \alpha , \ \ \ \beta \circ \epsilon \circ \beta = \beta.$$
Since $\alpha$ and $\beta$ are surjetive we get
$$ \alpha \circ \epsilon = Id_{\Gamma}, \ \ \ \beta \circ \epsilon = Id_{\Gamma}.$$
\item[(4)] $\ \ $ For each $g \in \Gamma$,
$$i\left(g\right) \cdot g = \epsilon \left(\alpha\left(g\right)\right) , \ \ \ g \cdot i\left(g\right) = \epsilon \left(\beta\left(g\right)\right).$$
Then,
$$ \alpha \circ i = \beta , \ \ \ \beta \circ i = \alpha.$$
\end{itemize}
These maps $\alpha, \beta, i,  \epsilon$ will be called \textit{structure maps}. In what follows, we will denote this groupoid by $ \Gamma \rightrightarrows M$.
\end{definition}
If $\Gamma$ is a groupoid over $M$, then $M$ is also denoted by $\Gamma_{\left(0\right)}$ and it is often identified with the set $\epsilon \left(M\right)$ of identity elements of $\Gamma$. $\Gamma$ is also denoted by $\Gamma_{\left(1\right)}$. The map $\left(\alpha , \beta\right) : \Gamma \rightarrow M \times M$ is called the \textit{anchor} of the groupoid.\\

\vspace{0.2cm}

Now, we define the morphisms in the category of groupoids.
\begin{definition}
\rm
If $\Gamma_{1} \rightrightarrows M_{1}$ and $\Gamma_{2} \rightrightarrows M_{2}$ are two groupoids then a morphism from $\Gamma_{1} \rightrightarrows M_{1}$ to $\Gamma_{2} \rightrightarrows M_{2}$ consists of two maps $\Phi : \Gamma_{1} \rightarrow \Gamma_{2}$ and $\phi : M_{1} \rightarrow M_{2}$ such that for any $g_{1} \in \Gamma_{1}$
\begin{equation}\label{1}
\alpha_{2} \left( \Phi \left(g_{1}\right)\right) = \phi \left(\alpha_{1} \left(g_{1} \right)\right), \ \ \ \ \ \ \ \beta_{2} \left( \Phi \left(g_{1}\right)\right) = \phi \left(\beta_{1} \left(g_{1} \right)\right),
\end{equation}
where $\alpha_{i}$ and $\beta_{i}$ are the source and the target map of $\Gamma_{i} \rightrightarrows M_{i}$ respectively, for $i=1,2$, and preserves the composition, i.e.,
$$\Phi \left( g_{1} \cdot h_{1} \right) = \Phi \left(g_{1}\right) \cdot \Phi \left(h_{1}\right), \ \forall \left(g_{1} , h_{1} \right) \in \left( \Gamma_{1}\right)_{\left(2\right)}.$$
We will denote this morphism by $\left(\Phi , \phi\right)$ or by $\Phi$ (because, using Eq. (\ref{1}), $\phi$ is completely determined by $\Phi$).
\end{definition}
Observe that, as a consequence, $\Phi$ preserves the identities, i.e., denoting by $\epsilon_{i}$ the section of identities of $\Gamma_{i} \rightrightarrows M_{i}$ for $i=1,2$, we have
$$\Phi \circ  \epsilon_{1} = \epsilon_{2} \circ \phi .$$
 
Using this definition we define a \textit{subgroupoid} of a groupoid $\Gamma \rightrightarrows M$ as a groupoid $\Gamma' \rightrightarrows M'$ such that $M' \subseteq M$, $\Gamma' \subseteq \Gamma$ and the corresponding inclusion map is a morphism of groupoids.

\begin{remark}
\rm
There is a more abstract way of defining a groupoid. A groupoid is a "small" category (the class of objects and the class of morphisms are sets) in which each morphism is invertible.\\
If $ \Gamma \rightrightarrows M$ is the groupoid, then $M$ is the set of objects and $\Gamma$ is the set of morphisms. In this sense, we can think about a groupoid as a set $M$ of objects and a set $\Gamma$ of invertible maps between objects of $M$. Then, for each map $g \in \Gamma$, $\alpha \left( g \right)$ is the domain of $g$, $\beta \left( g \right)$ is the codomain $g$ and $i \left( g \right)$ is the inverse of $g$. For all $x \in M$, $\epsilon \left( x \right)$ is the identity map at $x$ and, finally, the operation $\cdot$ can be thought as the composition of maps.\\
A groupoid morphism is a functor between these categories, which is a more natural definition.

\end{remark}
Now, we present the most basic examples of groupoids.

\begin{example}\label{2}
\rm
A group is a groupoid over a point. In fact, let $G$ be a group and $e$ the identity element of $G$. Then, $G \rightrightarrows \{e\}$ is a groupoid, where the operation of the groupoid, $\cdot$, is just the operation in $G$.
\end{example}

\begin{example}\label{3}

\rm
For any set $M$, we can consider the product space $ M \times M$. Then $M \times M$ has a groupoid structure over $M$ such that
$$ \left( x , y \right) \cdot \left( z , x \right) = \left( z , y \right),$$
for all $x,y,z \in M$. $M \times M \rightrightarrows M$ is said to be the \textit{pair groupoid of $M$}.\\
Note that, if $\Gamma \rightrightarrows M$ is an arbitrary groupoid over $M$, then the anchor $\left(\alpha , \beta\right) : \Gamma \rightarrow M \times M$ is a morphism from $\Gamma \rightrightarrows M$ to the pair groupoid of $M$.
\end{example}

Next, we introduce the notion of orbits and isotropy group.
\begin{definition}
\rm
Let $\Gamma \rightrightarrows M$ be a groupoid with $\alpha$ and $\beta$ the source map and target map, respectively. For each $x \in M$, we denote
$$\Gamma_{x}^{x} = \beta^{-1}\left(x\right) \cap \alpha^{-1}\left(x\right),$$
which is called the \textit{isotropy group of} $\Gamma$ at $x$. The set
$$\mathcal{O}\left(x\right) = \beta\left(\alpha^{-1}\left(x\right)\right) = \alpha\left(\beta^{-1}\left( x\right)\right),$$
is called the \textit{orbit} of $x$, or \textit{the orbit} of $\Gamma$ through $x$.\\
\indent{If $\mathcal{O}\left(x\right) = M$ for all $x \in M$, or equivalently $\left(\alpha,\beta\right) : \Gamma  \rightarrow M \times M$ is a surjective map, then the groupoid $\Gamma \rightrightarrows M$ is called \textit{transitive}.}\\
\indent{Furthermore, the preimages of the source map $\alpha$ of a groupoid are called $\alpha-$\textit{fibres}. Those of the target map $\beta$ are called $\beta-$\textit{fibres}. We will usually denote the $\alpha-$fibre (resp. $\beta-$fibre) at a point $x$ by $\Gamma_{x}$ (resp. $\Gamma^{x}$).}
\end{definition}

\begin{definition}\label{4}
\rm  
Let $\Gamma \rightrightarrows M$ be a groupoid with $\alpha$ and $\beta$ the source and target map, respectively. We may define the left translation on $g \in \Gamma$ as the map $L_{g} : \beta^{-1} \left( \alpha\left(g\right)\right) \rightarrow \beta^{-1} \left(\beta\left(g\right)\right)$, given by
$$ h \mapsto g \cdot h .$$
Similarly, we may define the right translation on $g$, $R_{g} : \alpha^{-1}\left(\beta\left(g\right)\right) \rightarrow \alpha^{-1} \left( \alpha \left(g\right)\right)$. 
\end{definition}
Note that,
\begin{equation}\label{5} 
Id_{\beta^{-1}\left(x\right)} = L_{\epsilon \left(x\right)}.
\end{equation}
So, for all $ g \in \Gamma $, the left (resp. right) translation on $g$, $L_{g}$ (resp. $R_{g}$), is a bijective map with inverse $L_{i\left(g\right)}$ (resp. $R_{i\left(g\right)}$), where $i : \Gamma \rightarrow \Gamma$ is the inverse map.\\
\indent{Different structures (topological and geometrical) can be imposed on groupoids, depending on the context we are dealing with. We are interested in a particular case, the so-called Lie groupoids.}
\begin{definition}
\rm
A \textit{Lie groupoid} is a groupoid $\Gamma \rightrightarrows M$ such that $\Gamma$ and $M$ are smooth manifolds, and all the structure maps are smooth. Furthermore, the source and the target maps are submersions.\\
A \textit{Lie groupoid morphism} is a groupoid morphism which is differentiable.\\
\end{definition}

\begin{definition}
\rm
Let $\Gamma \rightrightarrows M$ be a Lie groupoid. A \textit{Lie subgroupoid} of $\Gamma \rightrightarrows M$ is a Lie groupoid $\Gamma' \rightrightarrows M'$ such that $\Gamma' $ and $M'$ are submanifolds of $\Gamma$ and $M$, respectively; and the pair given by the inclusion maps $j_{\Gamma'} : \Gamma' \hookrightarrow \Gamma$ $j_{M'} : M' \hookrightarrow M$ become a morphism of Lie groupoids.
\end{definition}
Observe that, taking into account that $ \alpha \circ \epsilon = Id_{M} = \beta \circ \epsilon$, then $\epsilon$ is an injective immersion.\\
\indent{On the other hand, in the case of a Lie groupoid, $L_{g}$ (resp. $R_{g}$) is clearly a diffeomorphism for every $g \in \Gamma$.}\\

\begin{example}\label{6}
\rm
A Lie group is a Lie groupoid over a point. 
%%In fact, Let $G$ be a Lie group and $e$ the identity element of $G$. Then $\{e\}$ is a Lie subgroup of $G$. Finally, it is easy to prove that the groupoid of the example \ref{1} is a Lie groupoid.\\
%%We may consider the groupoid $G \rightrightarrows \{e\}$. The operation of the Lie groupoid $\cdot$ is the operation of the $G$. It is easy to show that $G \rightrightarrows \{e\}$ is a groupoid.
\end{example}

\begin{example}
\rm
Let $M$ be a manifold. The pair groupoid $M \times M \rightrightarrows M$ is a Lie groupoid.
\end{example}
Next, we will introduce an example which will be fundamental in this paper.
\begin{example}\label{7}
\rm
Let $M$ be a manifold, and denote by $\Pi^{1} \left(M,M\right)$ the set of all vector space isomorphisms $L_{x,y}: T_{x}M \rightarrow T_{y}M$ for $x,y \in M$ or, equivalently, the space of the $1-$jets of local diffeomorphisms on $M$. An element of $\Pi^{1} \left(M,M\right)$ will by denoted by $j^{1}_{x,y}\psi$, where $\psi$ is a local diffeomorphism from $M$ into $M$ such that $\psi \left( x \right) = y$.\\
\indent{$\Pi^{1} \left(M,M\right)$ can be seen as a transitive groupoid over $M$ where, for all $x,y \in M$ and $j^{1}_{x,y}\psi , j^{1}_{y,z}\varphi \in \Pi^{1} \left(M,M\right)$, we have}
\begin{itemize}
\item[(i)] $\ \ $ $\alpha\left(j^{1}_{x,y}\psi\right) = x$
\item[(ii)] $\ \ $ $\beta\left(j^{1}_{x,y}\psi\right) = y$
\item[(iii)] $\ \ $ $j^{1}_{y,z}\varphi \cdot j^{1}_{x,y}\psi = j^{1}_{x,z} \left( \varphi \circ \psi \right)$
\end{itemize}
This groupoid is called the $1-$\textit{jets groupoid on $M$}. In fact, let $\left(x^{i}\right)$ and $\left(y^{j}\right)$ be local coordinate systems on open sets $U, V \subseteq M$. Then, we can consider a local coordinate system on $\Pi^{1} \left(M,M\right)$ given by
\begin{equation}\label{8}
\Pi^{1}\left(U,V\right) : \left(x^{i} , y^{j}, y^{j}_{i}\right),
\end{equation}
where, for each $ j^{1}_{x,y} \psi \in \Pi^{1}\left(U,V\right)$
\begin{itemize}
\item $x^{i} \left(j^{1}_{x,y} \psi\right) = x^{i} \left(x\right)$.
\item $y^{j} \left(j^{1}_{x,y}\psi \right) = y^{j} \left( y\right)$.
\item $y^{j}_{i}\left( j^{1}_{x,y}\psi\right)  = \left. \dfrac{\partial \left(y^{j}\circ \psi\right)}{\partial x^{i} }  \right | _{x}$.
\end{itemize}
These local coordinates turn this groupoid into a transitive Lie groupoid.
\end{example}

\section{Characteristic distribution}
\label{sec:3}

This section is devoted to construct the so-called \textit{characteristic distribution}. This object arises from the need of working with a groupoid which does not have a structure of Lie groupoid. In fact, this object endows the groupoid of a kind of ``differentiable'' structure. For a detailed study of the characteristic distribution, see \cite{CHARDIST}.\\
%%%%Thus, this section will be divided into two steps: First, we will endow to a groupoid contained in a Lie groupoid with a ``differentiable" structure given by a integrable distribution called \textit{Characteristic Distribution}. As a second step, we will use the foliation which integrates the characteristic distribution to separate the groupoid into ``differentiable parts", i.e., Lie goupoids.\\
Let $ \Gamma \rightrightarrows M$ be a Lie groupoid and $\overline{\Gamma}$ be a subgroupoid of $\Gamma$ (not necessarily a Lie subgroupoid of $\Gamma$) over the same manifold $M$. We will denote by $\overline{\alpha}$, $\overline{\beta}$, $\overline{\epsilon}$ and $\overline{i}$ the restrictions of the structure maps of $\Gamma$ to $\overline{\Gamma}$ (see the diagram below).\\

\begin{center}
 \begin{tikzcd}[column sep=huge,row sep=huge]
\overline{\Gamma}\arrow[r, hook, "j"] \arrow[rd, shift right=0.5ex] \arrow[rd, shift left=0.5ex]&\Gamma \arrow[d, shift right=0.5ex] \arrow[d, shift left=0.5ex] \\
& M 
 \end{tikzcd}
\end{center}

\noindent{where $j$ is the inclusion map. Now, we can construct a distribution $A \overline{\Gamma}^{T}$ over the manifold $\Gamma$ in the following way,
$$ g \in \Gamma \mapsto A \overline{\Gamma}^{T}_{g} \leq T_{g} \Gamma,$$
such that $A \overline{\Gamma}^{T}_{g}$ is generated by the (local) left-invariant vector fields $\Theta \in \frak X_{loc} \left( \Gamma \right)$ whose flow at the identities is totally contained in $\overline{\Gamma}$, i.e.,
\begin{itemize}\label{26}
\item[(i)] $\ \Theta$ is tangent to the $\beta-$fibres, 
$$ \Theta \left( g \right) \in T_{g} \beta^{-1} \left( \beta \left( g \right) \right),$$
for all $g$ in the domain of $\Theta$.\\
\item[(ii)]  $\ \ \Theta$ is invariant by left translations,
$$ \Theta \left( g \right) = T_{\epsilon \left( \alpha \left( g \right) \right) } L_{g} \left( \Theta \left( \epsilon \left( \alpha \left( g \right) \right) \right) \right),$$
for all $g $ in the domain of $\Theta$.\\
\item[(iii)]$\ \ $ The (local) flow $\varphi^{\Theta}_{t}$ of $\Theta$ satisfies
$$\varphi^{\Theta}_{t} \left( \epsilon \left( x \right)\right) \in \overline{\Gamma}, $$
for all $x \in M$.\\
\end{itemize}
Notice that, the set of local vector fields on $\Gamma$ satisfying (i), (ii) and (iii) is not empty. In fact, the zero vector field fulfills these conditions. It is remarkable that condition (iii) is equivalent to the following,
\begin{itemize}
\item[(iii)'] $\ \ \ \ $ The (local) flow $\varphi^{\Theta}_{t}$ of $\Theta$ at $\overline{g}$ is totally contained in $\overline{\Gamma}$, for all $\overline{g} \in \overline{\Gamma}$.
\end{itemize}
Then, roughly speaking, $A \overline{\Gamma}^{T}$ is generated by the left-invariant vector fields whose flows cannot cross $\overline{\Gamma}$. The distribution $A \overline{\Gamma}^{T}$ is called the \textit{characteristic distribution of $\overline{\Gamma}$}.\\
For the sake of simplicity, we will denote the family of the vector fields which satisfy conditions (i), (ii) and (iii) by $\mathcal{C}$. The elements of $\mathcal{C}$ will be called \textit{admissible vector fields}.\\

By using the structure of groupoid of $\Gamma$ and $\overline{\Gamma}$ we can construct a smooth distribution $A \overline{\Gamma}^{\sharp}$ on $M$ and a generalized vector bundle $ A \overline{\Gamma}$ such that for each $x \in M$, the fibres are defined by
\begin{eqnarray*}
A \overline{\Gamma}_{x} &=&  A \overline{\Gamma}^{T}_{\epsilon \left( x \right)}\\
A \overline{\Gamma}^{\sharp} _{x}  &=& T_{\epsilon \left( x \right) } \alpha \left( A \overline{\Gamma}_{x} \right)
\end{eqnarray*}
Therefore, the following diagram is commutative
\begin{large}
\begin{center}
 \begin{tikzcd}[column sep=huge,row sep=huge]
\Gamma\arrow[r, "A \overline{\Gamma}^{T}"] &\mathcal{P} \left( T \Gamma \right) \arrow[d, "T\alpha"] \\
 M \arrow[u,"\epsilon"] \arrow[r,"A \overline{\Gamma}^{\sharp}"] \arrow[ru,dashrightarrow, "A \overline{\Gamma}"]&\mathcal{P} \left( T M \right)
 \end{tikzcd}
\end{center}
\end{large}

\vspace{35pt}
\noindent{where $\mathcal{P} \left( E \right)$ defines the power set of $E$.

The distribution $A \overline{\Gamma}^{\sharp}$ is called \textit{base-characteristic distribution of $\overline{\Gamma}$}. It is remarkable that both distributions are characterized by $A \overline{\Gamma}$ in the following way
$$ A \overline{\Gamma}^{T}_{g} = T_{\epsilon \left( \alpha \left( g \right) \right)} L_{g} \left( A \overline{\Gamma}_{ \alpha \left( g \right) } \right), \ \forall g \in \Gamma.$$

\indent{Summarizing, associated to $\overline{\Gamma}$, we have three mathematical objects $A \overline{\Gamma}$, $A \overline{\Gamma}^{T}$ and $A \overline{\Gamma}^{\sharp}$. Next, let us recall the importance of these objects.}\\

Consider a left-invariant vector field $\Theta$ on $\Gamma$ whose (local) flow $\varphi_{t}^{\Theta}$ at the identities is contained in $\overline{\Gamma}$. Then, the characteristic distribution $A \overline{\Gamma}^T$ is invariant by the flow $\varphi_{t}^{\Theta}$, i.e., for all $g \in \Gamma$ and $t$ in the domain of $\varphi_{g}^{\Theta}$ we have
\begin{equation}\label{9}
T_{g} \varphi_{t}^{\Theta} \left( A \overline{\Gamma}^{T}_{g} \right) = A \overline{\Gamma}^{T}_{\varphi_{t}^{\Theta} \left( g \right)}.
\end{equation}

In fact, this is an immediate consequence of that the composition of flows of left-invariant vector fiels in $\mathcal{C}$ results in a left-invariant flow which stays inside $\overline{\Gamma}$ for every $\overline{g} \in \overline{\Gamma}$.\\
Thus, $A \overline{\Gamma}^{T}$ is invariant by the generating family of vector fields $\mathcal{C}$. Now,
let us recall a classical theorem, due to Stefan \cite{PS} and Sussmann \cite{HJS}, which charaterizes the integrability of singular distributions.\\
\begin{theorem}[Stefan-Sussmann]\label{10}
Let $D$ be a smooth singular distribution on a smooth manifold $M$. Then the following three conditions are equivalent:
\begin{itemize}
\item[(a)] $\ \ $ $D$ is integrable.
\item[(b)] $\ $ $D$ is generated by a family $C$ of smooth vector fields, and is invariant with respect to $C$.
\item[(c)] $\ $ $D$ is the tangent distribution $D^{\mathcal{F}}$ of a smooth singular foliation $\mathcal{F}$.
\end{itemize}
\end{theorem}

Hence, there exists a foliation $\overline{\mathcal{F}}$ on $\Gamma$ which integrates $A \overline{\Gamma}^{T}$, i.e., at each point $g \in \Gamma$ the leaf $\overline{\mathcal{F}} \left( g \right)$ at $g$ satisfies that
$$ T_{g} \overline{\mathcal{F}} \left( g \right) = A \overline{\Gamma}^{T}_{g}.$$

The set of the leaves of $\overline{\mathcal{F}}$ at points in $\overline{\Gamma}$ is called the \textit{characteristic foliation of $\overline{\Gamma}$}. Note that, the characteristic foliation of $\overline{\Gamma}$ does not define a foliation on $\overline{\Gamma}$ because $\overline{\Gamma}$ is not, necessarily, a manifold.\\

We already have the following result.
\begin{theorem}\label{11}
Let $\Gamma \rightrightarrows M$ be a Lie groupoid and $\overline{\Gamma}$ be a subgroupoid of $\Gamma$ (not necessarily a Lie groupoid) over $M$. Then, there exists a foliation $\overline{\mathcal{F}}$ of $\Gamma$ such that $\overline{\Gamma}$ is a union of leaves of $\overline{\mathcal{F}}$.
\end{theorem}
In this way, $\overline{\Gamma}$ which is not a manifold has some kind of ``differentiable'' structure via the foliation $\overline{\mathcal{F}}$.\\
Let us highlight the following assertions
\begin{itemize}
\item[(i)]$\ $  For each $g \in \overline{\Gamma}$, then
$$\overline{\mathcal{F}} \left( g \right) \subseteq \overline{\Gamma}^{\beta \left( g \right)}.$$
\item[(ii)]$\ \ $  For each $g ,h \in \Gamma$ such that $\alpha \left( g \right) = \beta \left( h \right)$, we have
$$\overline{\mathcal{F}} \left( g \cdot h\right) = g \cdot \overline{\mathcal{F}} \left(  h\right).$$
\item[(iii)]$\ \ $  Let $\Theta \in \mathfrak{X}_{loc} \left( \Gamma \right)$ be a left-invariant vector field on $\Gamma$. Then, $\Theta \in \mathcal{C}$ if, and only if, 
\begin{equation}\label{12}
 \Theta_{|\overline{\mathcal{F}}\left( g \right)} \in \frak X \left( \overline{\mathcal{F}} \left( g \right) \right),
\end{equation}
for all $g $ in the domain of $\Theta$
\end{itemize}

The construction of the characteristic distribution imposes some condition of maximality, i.e., \textit{ $\overline{\beta}^{-1} \left( x \right)$ is a submanifold of $\Gamma$ for all $x \in M$ if, and only if, $\overline{\beta}^{-1} \left( x \right) = \overline{\mathcal{F}} \left( \epsilon \left( x \right) \right)$ for all $x \in M$.}\\
Analogously, the base-characteristic distribution $A  \overline{\Gamma}^{\sharp}$ is integrable. Its associated foliation $\mathcal{F}$ of $M$ will be called the \textit{base-characteristic foliation of $\overline{\Gamma}$}.\\

Next, the groupoid structure is used to endow the leaves of $\mathcal{F} \left( x \right)$ of a Lie groupoid structure. First, by using the foliated atlas associated to $\mathcal{F}$ and $\overline{\mathcal{F}}$ we can prove the following result:

\begin{proposition}\label{13}
Let $\Gamma \rightrightarrows M$ be a Lie groupoid and $\overline{\Gamma}$ be a subgroupoid of $\Gamma$ with $\overline{\mathcal{F}}$ and $\mathcal{F}$ the characteristic foliation and the base-characteristic foliation of $\overline{\Gamma}$, respectively. Then, for all $x \in M$, the mapping
$$\alpha_{| \overline{\mathcal{F}} \left( \epsilon \left( x \right) \right)} : \overline{\mathcal{F}} \left( \epsilon \left( x \right) \right) \rightarrow \mathcal{F} \left( x \right),$$
is a surjective submersion.
\end{proposition}

As an interesting consequence we have the next corollary.
\begin{corollary}
Let $\Gamma \rightrightarrows M$ be a Lie groupoid and $\overline{\Gamma}$ be a subgroupoid of $\Gamma$. Then, the manifolds $\overline{\mathcal{F}} \left( \epsilon \left( x \right) \right) \cap \alpha^{-1} \left( x \right)$ are Lie subgroups of $\Gamma_{x}^{x}$ for all $x \in M$.

\end{corollary}

Thus, let us now construct the algebraic structure of a groupoid over the leaves of $\mathcal{F}$. Denote the groupoid generated by $\overline{\mathcal{F}} \left( \epsilon \left( x \right) \right) $ by $\overline{\Gamma} \left( \mathcal{F} \left( x \right) \right)$.\\

Then, $\overline{\Gamma} \left( \mathcal{F} \left( x \right) \right)$ is constructed by the next steps: For all $g ,h\in \overline{\mathcal{F}} \left( \epsilon \left( x \right) \right) $, we have 
\begin{itemize}

\item[] If $\alpha \left( h \right) = \beta \left( g \right)$,
$$h \cdot g \in \overline{\Gamma} \left( \mathcal{F} \left( x \right) \right).$$
\item[] If $\alpha \left( h \right) = \alpha \left( g \right)$,
$$h \cdot g^{-1} \in \overline{\Gamma} \left( \mathcal{F} \left( x \right) \right).$$
\item[] If $\beta \left( h \right) = \beta \left( g \right)$,
$$h^{-1} \cdot g \in \overline{\Gamma} \left( \mathcal{F} \left( x \right) \right).$$
\end{itemize}

Equivalently, $\overline{\Gamma} \left( \mathcal{F} \left( x \right) \right)$ is the smallest transitive subgroupoid of $\overline{\Gamma}$ which contains $\overline{\mathcal{F}} \left( \epsilon \left( x \right) \right)$. Actually, we have that
\begin{equation}\label{15}
 \overline{\Gamma} \left( \mathcal{F} \left( x \right) \right) = \sqcup_{g \in \overline{\mathcal{F}} \left( \epsilon \left( x \right) \right)} \overline{\mathcal{F}} \left( \epsilon \left( \alpha \left( g \right) \right) \right),
\end{equation}
i.e., $\overline{\Gamma} \left( \mathcal{F} \left( x \right) \right)$ can be depicted as a disjoint union of fibres at the identities.\\
Observe that the $\beta-$fibre of this groupoid at a point $y \in \mathcal{F} \left( x \right)$ is given by $\overline{\mathcal{F}} \left( \epsilon \left( y \right) \right)$. Hence, the $\alpha-$fibre at $y$ is 
$$ \overline{\mathcal{F}}^{-1} \left( \epsilon \left( y\right) \right) = i \circ \overline{\mathcal{F}} \left( \epsilon \left( y \right) \right).$$
Furthermore, the Lie groups $\overline{\mathcal{F}} \left( \epsilon \left( y \right) \right) \cap \alpha^{-1} \left( y \right)$ are exactly the isotropy groups of $\overline{\Gamma}\left( \mathcal{F} \left( x \right) \right) $. All these results imply the following one (\cite{CHARDIST})
\begin{theorem}\label{16}
For each $x \in M$ there exists a transitive Lie subgroupoid $\overline{\Gamma} \left( \mathcal{F} \left( x \right) \right)$ of $\Gamma$ with base $\mathcal{F} \left( x \right)$.
\end{theorem}
So, we have divided the manifold $M$ into leaves $\mathcal{F} \left( x \right)$ which have a maximal structure of transitive Lie subgroupoids of $\Gamma$. In fact, $\overline{\Gamma}$ is a transitive Lie subgroupoid of $\Gamma$ if, and only if, $M = \mathcal{F} \left( x \right)$ and $\overline{\Gamma} =\overline{\Gamma} \left( \mathcal{F} \left( x \right) \right)$.
\begin{remark}
\rm
This construction of the characteristic distribution associated to a subgroupoid $\overline{\Gamma}$ of a Lie groupoid $\Gamma$ generalizes the known correspondence between Lie groupoids and Lie algebroids (see \cite{KMG}). Indeed, $A \overline{\Gamma}$ is the associated Lie algebroid to $\overline{\Gamma}$
 if $\overline{\Gamma}$ is a Lie subgroupoid of $\Gamma$.
%%%As another observation, given a subgroupoid $\overline{\Gamma}$ of a Lie groupoid $\Gamma$ we can still construct another integrable distribution over $\Gamma$ such that generates a foliation in which $\overline{\Gamma}$ is a union of leaves. This distribution will be generated by the vector fields on $\Gamma$ such that the flow at the point of $\overline{\Gamma}$ is totally contained in $\overline{\Gamma}$ (without the condition of tangency in the $\beta-$fibres). For each $x \in M$, the leaf of this foliation at $\epsilon \left( x \right)$ is $\overline{\Gamma} \left( \mathcal{F} \left( x \right) \right)$.

\end{remark}

\section{Material groupoid and Material Distribution}\label{18}

In this section we will apply the results of the section 3 to the case of continuum mechanics. First, let us fix the fundamental notions.\\
A \textit{body} $\mathcal{B}$ is a $3$-dimensional differentiable manifold which can be covered with just one chart. An embedding $\phi : \mathcal{B} \rightarrow \mathbb{R}^{3}$ is called a \textit{configuration of} $\mathcal{B}$ and its $1-$jet $j_{X,\phi \left(X\right)}^{1} \phi$ at $X \in \mathcal{B}$ is called an \textit{infinitesimal configuration at $X$}. We usually identify the body with any one of its configurations, say $\phi_{0}$, called \textit{reference configuration}. Given any arbitrary configuration $\phi$, the change of configurations $\kappa = \phi \circ \phi_{0}^{-1}$ is called a \textit{deformation}, and its $1-$jet $j_{\phi_{0}\left(X\right) , \phi \left(X\right)}^{1} \kappa$ is called an \textit{infinitesimal deformation at $\phi_{0}\left(X\right)$}.\\
In the case of simple bodies, the mechanical response of the material is characterized by one function $W$ which depends, at each point $X \in \mathcal{B}$, on the gradient of the deformations evaluated at the point. Thus, $W$ is defined as a differentiable map
$$ W : \Pi^{1} \left( \mathcal{B}, \mathcal{B}\right) \rightarrow V,$$
which does not depend on the final point with respect to the reference configuration, i.e., for all $X,Y,Z \in \mathcal{B}$
\begin{equation}\label{19}
 W \left( j_{X,Y}^{1} \phi\right) = W \left( j_{X,Z}^{1} \left( \phi_{0}^{-1} \circ \tau_{Z-Y} \circ \phi_{0} \circ \phi\right)\right), \ \forall j_{X,Y}^{1}\phi \in \Pi^{1} \left( \mathcal{B}, \mathcal{B}\right),
\end{equation}
where $V$ is a real vector space and $\tau_{v}$ is the translation map on $\mathbb{R}^{3}$ by the vector $v$. This map is called \textit{mechanical response}. There are other equivalent definitions (\cite{MELZA}, \cite{MELZASEG}, \cite{MEPMDLSEG} or \cite{VMJIMM}) of this function . We will use this definition for convenience.\\

%%%%by restriction any material submanifold $\mathcal{P}$ of a body $\mathcal{B}$ has structure of body in itself. In fact, let us consider a local diffeomorphism $\phi$ on $\mathcal{B}$ such that $\phi \left( \mathcal{P}\right) \subseteq \mathcal{P}$. Then, $\phi$ can be seen as a local diffeomorphism on $\mathcal{P}$. Conversely, restricting to suitable neighbourhoods, any local diffeomorphism $\phi$ on $\mathcal{P}$ can be extended to a local diffeomorphism $\tilde{\phi}$ on $\mathcal{B}$. So, the groupoid $\Pi^{1} \left( \mathcal{P} , \mathcal{P} \right)$ can be seen (not uniquely) as a subset of $ \Pi^{1} \left( \mathcal{B} , \mathcal{B} \right)$.\\

Now, consider a situation in which an open negihbourhood of a point $Y \in \mathcal{B}$ is diffeormorphic to an open neighbourhood of another point $X \in \mathcal{B}$ such that the diffeomorphism cannot be detected by a mechanical experiment. Then, roughly speaking, we will say that $Y$ and $X$ are made of the same material. In the case of this property is satisfied for any point in $\mathcal{B}$ we will say that the body is \textit{uniform}.

\begin{definition}
\rm
A body $\mathcal{B}$ is said to be \textit{uniform} if for each two points $X,Y \in \mathcal{B}$ there exists a local diffeomorphism $\psi$ from an open neighbourhood $U \subseteq \mathcal{B}$ of $X$ to an open neighbourhood $V \subseteq \mathcal{B}$ of $Y$ such that $\psi \left(X\right) =Y$ and
\begin{equation}\label{20}
W \left( j^{1}_{Y, \kappa \left(Y\right)} \kappa \cdot j^{1}_{X,Y} \psi \right) = W \left( j^{1}_{Y, \kappa \left(Y\right)} \kappa\right),
\end{equation}
for all infinitesimal deformation $j^{1}_{Y , \kappa \left(Y\right)} \kappa$. $j^{1}_{X,Y} \psi$ is called a \textit{material isomorphism}.
\end{definition}

Let us now consider the family of all material isomorphisms denoted by $\Omega \left( \mathcal{B} \right)$. It is a straighforward exercise to prove that $\Omega \left( \mathcal{B} \right)$ has a natural estructure of groupoid by using the composition of $1-$jets as the composition law of the groupoid. A material isomorphism from $X$ to $X$ is said to be a \textit{material symmetry}. We will denote the structure maps of the material groupoid $\Omega \left( \mathcal{B} \right)$ by $\overline{\alpha}$, $\overline{\beta}$, $\overline{\epsilon}$ and $\overline{i}$ which are, indeed, the restrictions of the corresponding ones on $\Pi^{1} \left(  \mathcal{B} , \mathcal{B} \right)$.\\
$\Omega \left( \mathcal{B} \right)$ is a subgoupoid of the Lie groupoid of the $1-$jets $\Pi^{1}\left( \mathcal{B} , \mathcal{B}\right)$. However, $\Omega \left( \mathcal{B} \right)$ is not necessarily a Lie subgroupoid of $\Pi^{1} \left( \mathcal{B} , \mathcal{B} \right)$ (see the examples below) and, hence, we are in the conditions of Section \ref{sec:3}.\\

Taking into account the continuity of the mechanical response $W$ we have that for any $X \in \mathcal{B}$ the group of material symmetries $\Omega \left( \mathcal{B} \right)_{X}^{X}$ is a closed subgroup of $\Pi^{1} \left(\mathcal{B} , \mathcal{B}\right)_{X}^{X}$. So, it follows this result.
\begin{proposition}\label{21}
Let $\mathcal{B}$ be a simple body. Then, for all $X \in \mathcal{B}$ the set of all material symmetries $\Omega \left( \mathcal{B} \right)_{X}^{X}$ is a Lie subgroup of $\Pi^{1} \left( \mathcal{B} , \mathcal{B}\right)_{X}^{X}$.
\end{proposition}
Notice that this result does not imply that $\Omega \left( \mathcal{B} \right)$ is a Lie subgroupoid of $\Pi^{1} \left( \mathcal{B} , \mathcal{B}\right)$. This is a consequence of that $\beta-$fibres of $\Omega \left( \mathcal{B} \right)$ could have different dimensions.\\

Now, let us express the uniformity as a known property of Lie groupoids.

\begin{proposition}
Let $\mathcal{B}$ be a body. $\mathcal{B}$ is uniform if, and only if, $\Omega \left( \mathcal{B}\right)$ is a transitive subgroupoid of $\Pi^{1} \left( \mathcal{B} , \mathcal{B}\right)$.
\end{proposition}

Next, we will consider another (slightly more restrictive) notion of uniformity.
\begin{definition}\label{28}
\rm
A body $\mathcal{B}$ is said to be \textit{smoothly uniform} if for each point $X \in \mathcal{B}$ there is an neighbourhood $U $ around $X$ such that for all $Y \in U$ and $j_{Y,X}^{1} \phi \in \Omega \left( \mathcal{B} \right)$ there exists a smooth field of material isomorphisms $P$ at $X$ from $\epsilon \left( X \right)$ to $j_{Y,X}^{1} \phi$.
\end{definition}

Observe that a smooth field of material isomorphisms $P$ at $X$ is just a (local) differentiable section of the restriction of $\overline{\alpha}$ to $\Omega^{X}\left( \mathcal{B}\right)$
$$ \overline{\alpha}^{X} :\Omega^{X}\left( \mathcal{B}\right) \rightarrow \mathcal{B}.$$
The existence of these smooth fields of material isomorphism can be equivalently expressed as $\overline{\alpha}^{X}$ is a surjective submersion. Immediately we prove that smooth uniformity implies uniformity.\\
It is obvious that $\mathcal{B}$ is smoothly uniform if, and only if, for each two points $X,Y \in \mathcal{B}$ there are two open neighbourhoods $U , V \subseteq \mathcal{B}$ of $X$ and $Y$ respectively and $P : U \times V \rightarrow \Omega \left( \mathcal{B} \right) \subseteq \Pi^{1} \left( \mathcal{B} , \mathcal{B} \right)$, a smooth section of the anchor map $\left( \overline{\alpha} , \overline{\beta} \right)$. When $X=Y$ we can assume that $U=V$ and $P$ is a morphism of groupoids over the identity map, i.e.,

$$ P \left( Z, T \right) = P \left(  R , T\right) P \left( Z, R \right), \ \forall \ T,R,Z \in U.$$ 

So, we have the following corollary of Proposition \ref{21}.

\begin{corollary}
Let $\mathcal{B}$ be a body. $\mathcal{B}$ is smoothly uniform if, and only if, $\Omega \left( \mathcal{B}\right)$ is a transitive Lie subgroupoid of $\Pi^{1} \left( \mathcal{B} , \mathcal{B}\right)$.
\end{corollary}

%%%\begin{proposition}
%%%Let $\mathcal{B}$ be a body. $\mathcal{B}$ is smoothly uniform if and only if $\Omega \left( \mathcal{B}\right)$ is a transitive Lie subgroupoid of $\Pi^{1} \left( \mathcal{B} , \mathcal{B}\right)$.
%%%\end{proposition}

\begin{remark}
\rm

Let $\Theta$ be an admissible left-invariant vector field on $\Pi^{1} \left( \mathcal{B} , \mathcal{B} \right)$ (see Section \ref{sec:3}), i.e., $\varphi^{\Theta}_{t}\left( \epsilon \left( X \right) \right) \in \Omega \left( \mathcal{B} \right)$ for all $X \in \mathcal{B}$. Then, for all $g \in \Pi^{1} \left( \mathcal{B} , \mathcal{B} \right)$, we have that
\begin{eqnarray*}
TW \left( \Theta \left( g \right)\right) &=& \left. \dfrac{\partial}{\partial t}\left(W \left(\varphi^{\Theta}_{t}\left( g \right) \right)\right) \right | _{0}\\
&=& \left. \dfrac{\partial}{\partial t}\left(W \left(g \cdot\varphi^{\Theta}_{t}\left( \epsilon \left( \alpha \left( g \right) \right) \right) \right)\right)  \right | _{0}\\
&=& \left. \dfrac{\partial}{\partial t}\left(W \left(g \right)\right)  \right | _{0}= 0.
\end{eqnarray*}
Therefore,
\begin{equation}\label{22}
TW \left( \Theta \right) = 0
\end{equation}
The converse is proved in the same way.\\
So, the characteristic distribution $A \Omega \left( \mathcal{B} \right)^{T}$ of the material groupoid is generated by the left-invariant vector fields on $\Pi^{1} \left( \mathcal{B} , \mathcal{B} \right)$ which are in the kernel of $TW$. This characteristic distribution will be called \textit{material distribution}. The base-characteristic distribution $A \Omega \left( \mathcal{B} \right)^{\sharp}$ will be called \textit{body-material distribution}. Let us recall that the left-invariant vector fields on $\Pi^{1} \left( \mathcal{B} , \mathcal{B} \right)$ which satisfy Eq. (\ref{22}) are called admissible vector fields and the family of these vector fields is denoted by $\mathcal{C}$.\\
\end{remark}
Denote by $\overline{\mathcal{F}} \left( \epsilon \left( X \right) \right)$ and $\mathcal{F} \left( X \right)$ the foliations associated to the material distribution and the body-material distribution respectively. For each $X \in \mathcal{B}$, we will denote the Lie groupoid $\Omega \left( \mathcal{B} \right)\left(\mathcal{F}\left( X \right)\right)$ by $\Omega \left( \mathcal{F} \left( X \right) \right)$.\\
\indent{Notice that, strictly speaking, in continuum mechanics a \textit{sub-body} of a body $\mathcal{B}$ is just an open submanifold of $\mathcal{B}$ but, here, the foliation $\mathcal{F}$ gives us submanifolds of different dimensions. So, we will consider a more general definition so that, a \textit{material submanifold (or generalized sub-body) of $\mathcal{B}$} is just a submanifold of $\mathcal{B}$. A generalized sub-body $\mathcal{P}$ inherits certain material structure from $\mathcal{B}$. In fact, we will measure the material response of a material submanifold $\mathcal{P}$ by restricting $W$ to the $1-$jets of local diffeomorphisms $\phi$ on $\mathcal{B}$ from $\mathcal{P}$ to $\mathcal{P}$. However, it easy to observe that a material submanifold of a body is not exactly a body. See \cite{MD} for a discussion on this subject.}

Then, as a corollary of Theorem \ref{16}, we have the following result.
\begin{theorem}
For all $X \in \mathcal{B}$, $\Omega \left( \mathcal{F} \left( X \right) \right)$ is a transitive Lie subgroupoid of $\Pi^{1} \left( \mathcal{B} , \mathcal{B} \right)$. Thus, any body $\mathcal{B}$ can be covered by a maximal foliation of smoothly uniform material submanifolds.
\end{theorem}
Notice that, in this case ``maximal'' means that any other foliation $\mathcal{G}$ by smoothly uniform material submanifolds is thinner than $\mathcal{F}$, i.e.,
$$ \mathcal{G} \left( X \right) \subseteq  \mathcal{F} \left( X \right)  , \ \forall X \in  \mathcal{B}.$$

\begin{remark}
\rm
Just imagine that there is, at least, a $1-$jet $g \in \Omega^{X} \left( \mathcal{B} \right)$ for some $X \in \mathcal{B}$ such that
$$ g \notin \overline{\mathcal{F}} \left( \epsilon \left( X \right) \right).$$
Then, we are not including $g$ inside any of the transitivie Lie subgroupoids $\Omega \left( \mathcal{F} \left( X \right) \right)$. Thus, these material isomorphisms are being discarded.\\
Nevertheless
\begin{equation}\label{23}
\overline{\mathcal{F}} \left( g \right) = g \cdot \overline{\mathcal{F}}\left( \epsilon \left( \alpha \left( g \right) \right) \right),
\end{equation}
and, indeed, $ \overline{\mathcal{F}}\left( \epsilon \left( \alpha \left( g \right) \right) \right)$ is contained in $\Omega \left( \mathcal{F} \left( \alpha \left( g \right) \right)\right)$, i.e., using Eq. (\ref{23}), we can reconstruct $\overline{\mathcal{F}} \left( g \right)$.
\hfill
\end{remark}

%%%%%%To summarize, we have introduced a differentiable structure (the characteristic distribution) to study any groupoid $\overline{\Gamma}$ contained in a Lie groupoid without assuming that $\overline{\Gamma}$ is a Lie groupoid. Then, we have used this development for the set of material isomorphisms (the material groupoid). Using this object, we have been able to prove that any material body can be divided into smoothly uniform material submanifolds. So, roughly speaking, any body can be studied as the union of uniform bodies which have differentiable material groupoids and this union has an useful particular structure (foliated charts). Finally, we could use the material distribution to study if the material groupoid is a Lie subgroupoid of $\Pi^{1} \left( \mathcal{B} , \mathcal{B} \right)$ or not (see Remark \ref{22}), i.e., to study the differentiability of the set of material isomorphisms.\\

Finally, using the body-material distribution, we will be able to define a more general notion of smooth uniformity. This notion was introduced in \cite{MGEOEPS}. We will end up using the foliation by uniform subbodies to interpret it over the material groupoid.

\begin{definition}
\rm
Let be a body $\mathcal{B}$ and a body point $X \in \mathcal{B}$. Then, $\mathcal{B}$ is said to be \textit{uniform of grade $p$ at $X$} if $A\Omega \left( \mathcal{B} \right)^{\sharp} _{X}$ has dimension $p$. $\mathcal{B}$ is \textit{uniform of grade $p$} if it is uniform of grade $p$ at all the points.
\end{definition}

Note that, smooth uniformity is a particular case of graded uniformity. In fact, $\mathcal{B}$ is smoothly uniform if, and only if, $\mathcal{B}$ is uniform of grade $n$. Equivalently, $\mathcal{B}$ is uniform of grade $3$ if, and only if, $A \Omega \left( \mathcal{B} \right)^{\sharp}_{X}$ has dimension $3$ for all $X \in \mathcal{B}$, i.e., there exists just one leaf of the material foliation equal to $\mathcal{B}$. Hence, the material groupoid $\Omega \left( \mathcal{B} \right)$ is a Lie subgroupoid of $\Pi^{1}\left( \mathcal{B} , \mathcal{B} \right)$ whose $\overline{\beta}-$fibres integrate the material distribution.

\begin{corollary}
Let be a body $\mathcal{B}$ and a body point $X \in \mathcal{B}$. $\mathcal{B}$ is uniform of grade $p$ at $X$ if, and only if, the uniform leaf $\mathcal{F} \left( X \right)$ at $X$ has dimension $p$.
\end{corollary}

\begin{corollary}
Let $\mathcal{B}$ be a body. $\mathcal{B}$ is uniform of grade $p$ if, and only if, the body-material foliation is regular of rank $p$.
\end{corollary}
 
It is important to highlight that the body-material foliation has certain condition of maximality. In fact, suppose that there exists another foliation $ \mathcal{G}$ of $\mathcal{B}$ by smoothly uniform material submanifolds. Then, for all $X \in \mathcal{B}$ we have that
$$\mathcal{G} \left( X \right) \subseteq \mathcal{F} \left( X \right), \ \forall X \in \mathcal{B}.$$
So, we have the following results:
\begin{corollary}
Let be a body $\mathcal{B}$ and a body point $X \in \mathcal{B}$. $\mathcal{B}$ is uniform of grade greater or equal to $p$ at $X$ if, and only if, there exists a foliation $ \mathcal{G}$ of $\mathcal{B}$ by smoothly uniform submanifolds such that the leaf $\mathcal{G} \left( X \right)$ at $X$ has dimension greater or equal to $p$.
\end{corollary}

\begin{corollary}
Let $\mathcal{B}$ be a body. $\mathcal{B}$ is uniform of grade $p$ if, and only if, the body can be foliated by smoothly uniform material submanifold of dimension $p$.
\end{corollary}

\section{Homogeneity}
%%%%As originally introduced by Noll \cite{}, a precondition for a body to be homogeneous is that it must be materially uniform. In the putative physical picture, however, if homogeneity is to be interpreted as the absence of defects, it is conceivable that in some sense a non-uniform body may be defect-free and that, if it were not, a measure of defect density may be obtainable. A generalization of this kind for certain classes of FGMs has been presented in \cite{}, where the notion of material isomorphism was greatly relaxed by declaring points to be isomorphic if they just have conjugate symmetry groups. For $p-$uniform bodies of grades $1$ and $2$, which are indeed FGMs, stricter criteria of homogeneity can be established without thus relaxing the definition of material isomorphism.\\

This section is devoted to deal we the definition of homogeneity. As we already know, a body is uniform if the function $W$ does not depend on the point $X$. In addition, a body is said to be \textit{homogeneous} if we can choose a global section of the material groupoid which is constant on the body, more precisely:

\begin{definition}\label{29}
\rm
A body $\mathcal{B}$ is said to be \textit{homogeneous} if it admits a global configuration $\psi$ which induces a global section of $\left(\alpha , \beta\right)$ in $\Omega \left( \mathcal{B}\right)$, $P$, i.e., for each $X,Y \in \mathcal{B}$
$$ P\left(X,Y\right) = j^{1}_{X,Y} \left(\psi^{-1} \circ \tau_{\psi\left(Y\right) - \psi \left(X\right)} \circ \psi\right),$$
where $\tau_{\psi\left(Y\right) - \psi \left(X\right)}: \mathbb{R}^{3} \rightarrow \mathbb{R}^{3}$ denotes the translation on $\mathbb{R}^{3}$ by the vector $\psi\left(Y\right) - \psi \left(X\right)$. $\mathcal{B}$ is said to be \textit{locally homogeneous} if there exists a covering of $\mathcal{B}$ by homogeneous open sets. $\mathcal{B}$ is said to be \textit{(locally) inhomogeneous} if it is not (locally) homogeneous.
\end{definition}

Notice that local homogeneity is clearly more restrictive than smooth uniformity. In fact, in this case, the smooth fields of material isomorphisms (see Definition \ref{28}) are induced by particular (local) configurations. However, in a purely intuitive picture, homogeneity can be interpreted as the absence of defects. So, it makes sense to develop some kind of homogeneity for non-uniform material which measures the absence of defects and generalizes the known one. In the literature we can already find some partial answer of this question (\cite{FGMA,FGM2} for FGM's and \cite{EPST,MGEOEPS} for laminated and bundle materials).

Recall that the material distributions are characterized by the commutativity of the following diagram

\begin{large}
\begin{center}
 \begin{tikzcd}[column sep=huge,row sep=huge]
\Pi^{1}\left( \mathcal{B} , \mathcal{B} \right) \arrow[r, "A \Omega \left(\mathcal{B} \right)^{T}"] & P \left( T \Pi^{1}\left( \mathcal{B} , \mathcal{B} \right)  \right) \arrow[d, "T\alpha"] \\
 \mathcal{B} \arrow[u,"\epsilon"] \arrow[r,"A  \Omega \left(\mathcal{B} \right)^{\sharp}"] \arrow[ru,dashrightarrow]&P \left( T \mathcal{B} \right)
 \end{tikzcd}
\end{center}
\end{large}
As we have proved in the previous section, the body-material foliation $\mathcal{F}$ divides the body into smoothly uniform components.\\
Let us now give the intuition behind the definition of homogeneity of a non-uniform body. A non-uniform body will be \textit{(locally) homogeneous} whether each smoothly uniform material submanifold $\mathcal{F} \left( X\right)$ is (locally) homogeneous and all the uniform material submanifolds can be straightened at the same time.\\
Thus, we need to clarify what we understand by homogeneity of submanifolds of $\mathcal{B}$.
\begin{definition}
Let $\mathcal{B}$ be a simple body and $\mathcal{N}$ be a submanifold of $\mathcal{B}$. $\mathcal{N}$ is said to be \textit{homogeneous} if, and only if, for all point $X \in \mathcal{N}$ there exists a local configuration $\psi$ of $\mathcal{B}$ on an open subset $ U \subseteq \mathcal{B}$, with $\mathcal{N} \subseteq U$, which satisfies that

$$ j_{Y,Z}^{1} \left( \psi^{-1} \circ \tau_{ \psi\left( Z \right) - \psi \left( Y\right)} \circ \psi \right),$$
is a material isomorphism for all $Y,Z \in U \cap \mathcal{N}$. We will say that $\mathcal{N}$ is \textit{locally homogeneous} if there exists a covering of $\mathcal{N}$ by open subsets $U_{a}$ of $\mathcal{B}$ such that $U_{a} \cap \mathcal{N}$ are homogeneous submanifolds of $\mathcal{B}$. $\mathcal{N}$ is said to be \textit{(locally) inhomogeneous} if it is not (locally) homogeneous.
\end{definition}

Notice that, the definitions of homogeneity and local homogeneity for smoothly uniform materials (Definition \ref{29}) are generalized by this one whether $\mathcal{N}=\mathcal{B}$ or $\mathcal{N}$ is just an open subset of $\mathcal{B}$.\\

Now, taking into account that $\mathcal{F}=\{  \mathcal{F} \left( X \right) \}_{X \in \mathcal{B} }$ is a foliation, there is a kind of compatible atlas which are called \textit{foliated atlas}. In fact, $\{\left(\left(X^{I}_{a}\right), U_{a} \right)\}_{a}$ is a foliated atlas of $\mathcal{B}$ associated to $\mathcal{F}$ whether for each $X \in U_{a} \subseteq \mathcal{B}$ we have that $U_{a}:= \{ - \epsilon < X^{1}_{a} < \epsilon , \hdots , - \epsilon < X^{3}_{a} < \epsilon \}$ for some $\epsilon > 0$, such that the $k-$dimensional disk $\{ X^{k+1}_{a}= \hdots = X^{3}_{a} = 0\}$ coincides with the path-connected component of the intersection of $\mathcal{F}\left( X \right)$ with $U_{a}$ which contains $X$, and each $k-$dimensional disk $\{ X^{k+1}_{a} = c_{k+1} , \hdots , X^{3}_{a} = c_{3} \}$, where $c_{k+1}, \hdots , c_{3}$ are constants, is wholly contained in some leaf of $\mathcal{F}$. Intuitively, this atlas straightens (locally) the partition $\mathcal{F}$ of $\mathcal{B}$.\\
The existence of these kind of atlas and the maximality condition over the smoothly uniform material submanifolds $\mathcal{F} \left( X \right)$ induces us to give the following definition.\\

\begin{definition}\label{31}
Let $\mathcal{B}$ be a simple body. $\mathcal{B}$ is said to be \textit{locally homogeneous} if, and only if, for all point $X \in \mathcal{B}$ there exists a local configuration $\psi$ of $\mathcal{B},$ with $X \in U$, which is a foliated chart and it satisfies that
$$ j_{Y,Z}^{1} \left( \psi^{-1} \circ \tau_{ \psi\left( Z \right) - \psi \left( Y\right)} \circ \psi \right),$$
is a material isomorphism for all $Z \in U \cap \mathcal{F}\left( Y \right)$. We will say that $\mathcal{B}$ is homogeneous if $U=\mathcal{B}$. $\mathcal{B}$ is said to be \textit{(locally) inhomogeneous} if it is not (locally) homogeneous.
\end{definition}
It is remarkable that, as we had said above, all the uniform leaves $\mathcal{F} \left( X \right)$ of an homogeneous body are homogeneous. Therefore, the definition of homogeneity for a smoothly uniform body coincides with Definition \ref{29}. Notice also that, the condition of all the leaves $\mathcal{F} \left( X \right)$ are homogeneous is not enough in order to have the homogeneity of the body $\mathcal{B}$ because there is also a condition of compatibility with the foliation structure of $\mathcal{F}$.\\

Let us recall a result given in \cite{MELZA} (see also \cite{MELZASEG} or \cite{CCWANSEG}) which characterizes the homogeneity by using $G-$structures.\\
Denote by $F\mathcal{B}$ the frame bundle of $\mathcal{B}$. An element of $F \mathcal{B}$ is called a linear frame at a point $X \in \mathcal{B}$ and it is simply a $1-$jet of a local diffeomorphism $f : \mathbb{R}^{3} \rightarrow \mathcal{B}$ at $0$ with $f \left( 0 \right) = X$. Then, the structure group of $F \mathcal{B}$ is the group of $3 \times 3-$regular matrices in $\mathbb{R}$, $Gl \left( 3 , \mathbb{R} \right)$.\\
A \textit{$G-$structure over} $\mathcal{B}$, $\omega_{G}\left(\mathcal{B}\right)$, is a reduced subbundle of $F\mathcal{B}$ with structure group $G$ a Lie subgroup of $Gl \left(3 , \mathbb{R}\right)$ (a good reference about frame bundles in \cite{DIGFMCORD}).\\
So, fix  $\overline{g}_{0} $ be a frame at $Z \in \mathcal{B}$. Then, assuming that $\mathcal{B}$ is smoothly uniform, the set
$$ \Omega \left( \mathcal{B} \right)_{Z} \cdot \overline{g}_{0} := \{ \overline{g} \cdot \overline{g}_{0} \ : \ \overline{g} \in \Omega \left( \mathcal{B} \right)_{Z} \},$$
where $\cdot$ defines the composition of $1-$jets, is a $\Omega \left( \mathcal{B} \right)_{Z}^{Z}-$structure over $\mathcal{B}$.

\begin{proposition}\label{24}
Let be a frame $\overline{g}_{0} \in F \mathcal{B}$. If $\mathcal{B}$ is homogeneous then the $G-$structure given by $\Omega \left( \mathcal{B} \right) \cdot \overline{g}_{0}$ is integrable. Conversely, $\Omega \left( \mathcal{B} \right) \cdot \overline{g}_{0}$ is integrable implies that $\mathcal{B}$ is locally homogeneous.
\end{proposition}

Thus, the next step will be to give a similar result for this generalized homogeneity. Because of the lack of uniformity we have to use groupoids instead of $G-$structures.\\

Let $\mathbb{S} := \{ \mathbb{S} \left( x \right) \ : \ x \in \mathbb{R}^{n} \}$ be a canonical foliation of $\mathbb{R}^{n}$, i.e., for all $x = \left( x^{1}, \hdots , x^{n} \right)\in \mathbb{R}^{n}$ the leaf $\mathbb{S}  \left(x \right)$ at $x$
$$ \mathbb{S} \left( x \right) := \{ \left( y^{1}, \hdots, y^{p}, x^{p+1}, \hdots, x^{n}\right) \ : \ y^{i} \in \mathbb{R}, \ i=1, \hdots, p \},$$
for some $1 \leq p \leq n$.\\
Notice that for any foliation $\mathcal{G}$ on a manifold $Q$ there exists a map
$$ p_{\mathcal{G}} : Q \rightarrow \{ 0, \hdots, dim\left(Q \right) \},$$
such that for all $x \in Q$ 
$$p_{\mathcal{G}} \left( x \right) = dim \left( \mathcal{G} \left( x \right) \right).$$
$p_{\mathcal{G}}$ will be called \textit{grade of $\mathcal{G}$}. $\mathcal{G}$ is a \textit{regular foliation} if, and only if, the grade of $\mathcal{G}$ is constant.\\
It is important to remark that in the case of $\mathbb{S}$ the grade $p_{\mathbb{S}}$ characterizes the foliation $\mathbb{S}$. Thus, as an abuse of notation, we could say that the map $p_{\mathbb{S}}$ is the foliation.\\

Let $\mathbb{S}$ be a canonical foliation of $\mathbb{R}^{n}$ with grade $p= p_{\mathbb{S}}$. Thus, as a generalization of the frame bundle of $\mathbb{R}^{n}$ we define the \textit{$p-$graded frame groupoid} as the following subgroupoid of $\Pi^{1} \left( \mathbb{R}^{n} , \mathbb{R}^{n} \right)$,
$$ \Pi^{1}_{p} \left( \mathbb{R}^{n} , \mathbb{R}^{n} \right) = \{ j^{1}_{x,y} \psi \in \Pi^{1} \left( \mathbb{R}^{n} , \mathbb{R}^{n} \right) \ : \ y \in \mathbb{S} \left( x \right) \}.$$
Notice that the restriction of $ \Pi^{1}_{p} \left( \mathbb{R}^{n} , \mathbb{R}^{n} \right)$ to any leaf $\mathbb{S} \left( x \right)$ is a transtive Lie subgroupoid of $ \Pi^{1} \left( \mathbb{R}^{n} , \mathbb{R}^{n} \right)$ with all the isotropy groups isomorphic to $Gl \left( n , \mathbb{R} \right)$. However, the groupoid $ \Pi^{1}_{p} \left( \mathbb{R}^{n} , \mathbb{R}^{n} \right)$ is not necessarily a Lie subgroupoid of $ \Pi^{1} \left( \mathbb{R}^{n} , \mathbb{R}^{n} \right)$. In fact, $ \Pi^{1}_{p} \left( \mathbb{R}^{n} , \mathbb{R}^{n} \right)$ is a Lie subgroupoid of $ \Pi^{1} \left( \mathbb{R}^{n} , \mathbb{R}^{n} \right)$ if, and only if, $\mathbb{S}$ is regular foliation.\\
A \textit{standard flat $G-$reduction of grade $p$} is a subgroupoid $ \Pi^{1}_{G,p} \left( \mathbb{R}^{n} , \mathbb{R}^{n} \right)$ of \linebreak $ \Pi^{1}_{p} \left( \mathbb{R}^{n} , \mathbb{R}^{n} \right)$ such that the restrictions $ \Pi^{1}_{G,p} \left( \mathbb{S} \left( x\right) , \mathbb{S} \left( x\right) \right)$ to the leaves $\mathbb{S} \left( x \right)$ are transitive Lie subgroupoids of $ \Pi^{1} \left( \mathbb{R}^{n} , \mathbb{R}^{n} \right)$ on the leaf $\mathbb{S} \left( x \right)$. It is remarkble that in this case all the isotropy groups of $\Pi^{1}_{G,p}\left( \mathbb{S} \left( x\right) ,  \mathbb{S} \left( x\right) \right)$ are conjugated.\\
Clearly, all the structures introduced in this section can be restricted to any open subset of $\mathbb{R}^{n}$.\\
Let $\psi : U \rightarrow \overline{U}$ be a (local) configuration on $U \subseteq \mathcal{B}$. Then, $\psi$ induces a Lie groupoids isomorphism,
$$
\begin{array}{rccl}
\Pi \psi : &  \Pi^{1} \left( U,U \right) & \rightarrow & \Pi^{1} \left( \overline{U}, \overline{U} \right)\\
& j_{X,Y}^{1}\phi &\mapsto & j_{\psi\left( X \right), \psi \left( Y \right)}^{1}\left( \psi \circ \phi \circ \psi^{-1}\right)
\end{array}
$$
\begin{proposition}
Let $\mathcal{B}$ be a simple body. If $\mathcal{B}$ is homogeneous the material groupoid is isomorphic (via a global configuration) to a standard flat $G-$reduction. Conversely, the material groupoid is isomorphic (via a local configuration) to a standard flat $G-$reduction implies that $\mathcal{B}$ is locally homogeneous.
\end{proposition}

Notice that, in the context of principal bundles, a $G-$structure is integrable if, and only if, there exist a local configuration which induces an isomorphism from the $G-$structure to a standard flat $G-$structure.\\

Finally, we will use the material distribution to give another characterization of homogeneity.\\
Let $\mathcal{B}$ be a homogeneous body with $\psi = \left( X^{I} \right)$ as an (local) homogeneous configuration. Then, by using that $\psi$ is a foliated chart we have that the partial derivaties are tangent to $A \Omega \left( \mathcal{B} \right)^{\sharp}$, i.e., for each $ X \in U$
$$ \dfrac{\partial}{\partial {X^{L}}_{|X}} \in A \Omega \left( \mathcal{B} \right)^{\sharp}_{X},$$
for all $1 \leq L \leq dim \left( \mathcal{F} \left( X \right) \right)=K$. Thus, there are local functions $\Lambda^{J}_{I,L}$ such that for each $L \leq K$ the (local) left-invariant vector field on $\Pi^{1} \left( \mathcal{B} , \mathcal{B} \right)$ given by
$$  \dfrac{\partial}{\partial X^{L}} + \Lambda^{J}_{I,L} \dfrac{\partial}{\partial {Y^{J}_{I}}},$$
are tangent to $A \Omega \left( \mathcal{B} \right)^{T}$ where $\left( X^{I}, Y^{J}, {Y^{J}_{I}} \right)$ are the induced coordinates of $\left( X^{I} \right) $ in $\Pi^{1} \left( \mathcal{B} , \mathcal{B} \right)$. Equivalently, the local functions $\Lambda^{J}_{I,L}$ satisfy that
$$   \dfrac{\partial W}{\partial X^{L}} + \Lambda^{J}_{I,L} \dfrac{\partial W}{\partial {Y^{J}_{I}}} = 0,$$
for all $1 \leq L \leq K$. Next, using that for each two points $X,Y \in U$ the $1-$jet $ j_{X,Y}^{1} \left( \psi^{-1} \circ \tau_{ \psi\left( Y \right) - \psi \left( X\right)} \circ \psi \right)$ is a material isomorphism we can choose $\Lambda^{J}_{I,L} = 0$.
\begin{proposition}\label{30}
Let $\mathcal{B}$ be a simple body. $\mathcal{B}$ is homogeneous if, and only if, for each $X\in \mathcal{B}$ there exists a local chart $\left( X^{I} \right)$ on $\mathcal{B}$ at $X$ such that,
\begin{equation}\label{25}
\dfrac{\partial W}{\partial X^{L}} = 0,
\end{equation}
for all $L\leq dim \left(\mathcal{F} \left( X \right)\right)$.
\end{proposition}
Notice that Eq. (\ref{25}) implies that the partial derivatives of the coordinates $\left( X^{I} \right)$ until $dim \left(\mathcal{F} \left( X \right)\right)$ are tangent to the material distribution and, therefore, the coordinates are foliated. So, Eq. (\ref{25}) gives us an apparently more straightforward way to express this general homogeneity.
\section{Examples}

We will devote this section to study the notion of homogeneity given in Definition \ref{31} for non-uniform material. In particular, we will present an example of homogeneous non-uniform material body (\textbf{Example 1}) and an example of a class of non-uniform materials where we can find inhomogeneous non-uniform materials (\textbf{Example 2}).

\subsection*{Example 1}\label{32}

Let $\mathcal{B}$ be a simple material in which there exists a reference configuration $\psi_{0}$ from $\mathcal{B}$ to the $3-$dimensional open cube $\mathcal{B}_{0} = \left(-1,1\right)^{3}$ in $\mathbb{R}^{3}$ that induces the following mechanical response
$$
\begin{array}{rccl}
W: & \Pi^{1}\left( \mathcal{B}_{0} , \mathcal{B}_{0} \right) \ & \rightarrow & \mathfrak{gl} \left( 3, \mathbb{R}\right)\\
& j_{X,Y}^{1}\phi &\mapsto & f \left(X^{1} \right) \left( F^{T}\cdot F - I \right),\\
\end{array}
$$
such that\\

$
f \left( X^{1}\right)= \left\{ \begin{array}{lcc}
            1 &   if  & X^{1} \leq 0 \\
             \\ 1 + e^{-\dfrac{1}{X^{1}}} &  if & X^{1}> 0 
             \end{array}
   \right.
$\\\\

\noindent{where $\mathfrak{gl} \left( 3, \mathbb{R}\right)$ is the algebra of matrices, $F$ is the Jacobian matrix of $\phi$ at $X$ respect to the canonical basis of $\mathbb{R}^{3}$ and $I$ is the identity matrix. Here, the (global) canonical coordinates of $\mathbb{R}^{3}$ are denoted by $\left( X^{i} \right)$ and $X = \left( X^{1},X^{2},X^{3} \right)$ respect to these coordinates.}\\
Notice that $f$ is constant until $0$ and strictly growing from $0$. Immediately, one can realize that $\mathcal{B}_{0}$ is not uniform. In fact, there are no material isomorphisms joining any two points $\left(X^{1},X^{2},X^{3}\right)$ and $\left( Y^{1},Y^{2},Y^{3}\right)$ such that
$$ f \left( X^{1} \right) \neq f \left( Y^{1}\right).$$
So, let us study the derivative of $W$ in order to find the grades of uniformity of the points of the body $\mathcal{B}_{0}$. The grades of uniformity for this example were first studied in \cite{MD}.\\

\begin{itemize}
\item[] $\dfrac{\partial W}{\partial F^{l}_{m}}  =  f \left( X^{1}\right) \left[ \left(\delta^{\left(k,j\right)}_{\left( l,m \right)} F^{k}_{i}\right)^{j}_{i} + \left(  F^{k}_{j} \delta^{\left( k,i \right)}_{\left( l,m \right)} \right)^{j}_{i}\right]$\\\\
\item[] $\dfrac{\partial W}{\partial X^{1}}  =  \dfrac{\partial f}{\partial X^{1}} \left(  F^{k}_{j}F^{k}_{i} \right)^{j}_{i}$\\\\
\item[] $\dfrac{\partial W}{\partial X^{i}}  = 0, \ i \geq 2.$
\end{itemize}
\noindent
Hence, we are looking for left-invariant (local) vector fields $\Theta$ on $\Pi^{1} \left( \mathcal{B}_{0} , \mathcal{B}_{0} \right)$ satisfying 
\begin{equation}\label{aa}
\Theta \left( W \right) = 0.
\end{equation}
Let $\left( X^{i},Y^{i},Y^{j}_{i} \right)$ be the induced coordinates of $\left( X^{i} \right)$ on $\Pi^{1} \left( \mathcal{B}_{0} , \mathcal{B}_{0} \right)$. Then, $\Theta$ can be expressed as follows,
$$\Theta \left( X^{i} , Y^{j} , F^{j}_{i} \right) = \left( \left( X^{i} , Y^{j} , F^{j}_{i} \right)  , \delta X^{i} ,0,  F^{j}_{l} \delta P^{l}_{i} \right).$$
Hence, $\Theta$ satisfies Eq. (\ref{aa}) if, and only if,
\begin{eqnarray*}\label{bb}
\Theta \left( W \right) &=& f \left( X^{1}\right) \left(F^{k}_{l}\delta P^{l}_{j} F^{k}_{i} + F^{k}_{j} F^{k}_{l}\delta P^{l}_{i}\right)^{j}_{i} \\
 &+&\delta X^{1} \dfrac{\partial f}{\partial X^{1}} \left(  F^{k}_{j}F^{k}_{i} - \delta^{j}_{i} \right)^{j}_{i} = 0.
\end{eqnarray*}

Let us focus first in the open given by the restriction $X^{1} < 0$. Then, Eq. (\ref{bb}), turns into the following,
\begin{equation}\label{ee}
F^{k}_{l}\delta P^{l}_{j} F^{k}_{i} + F^{k}_{j} F^{k}_{l}\delta P^{l}_{i} = 0, \ \forall i,j.
\end{equation}
for every Jacobian matrix $F = \left( F^{j}_{i}\right)$ of a local diffeomorphism $\phi$ on $\mathcal{B}_{0}$. Equivalently,
\begin{equation}
\delta P^{T} \cdot F^{T}\cdot F =- F^{T}\cdot F \cdot \delta P ,
\end{equation}
where $\delta P = \left( \delta P^{j}_{i}\right)$. Hence, $\delta P$ is a skew-symmetric matrix. So, for any family of local functions $\{ f^{i} , \delta P^{j}_{i}\}$ on the open restriction $\{X^{1} < 0 \}$ of the body $\mathcal{B}_{0}$ such that $\delta P = \left(\delta P^{j}_{i}\right)$ is an skew-symmetric matrix generates a vector field
$$\Theta \left( X^{i} , Y^{j} , F^{j}_{i} \right) = \left( \left( X^{i} , Y^{j} , F^{j}_{i} \right)  , f^{i} ,0,  F^{j}_{l} \delta P^{l}_{i} \right),$$
which satisfies Eq. (\ref{aa}). Therefore, the sub-body $\left(-1,1\right)^{3} \cap \{X^{1} < 0 \}$ is uniform.\\
Next we will study the open subset of $\mathcal{B}_{0}$ such that $X^{1} > 0$. Then, Eq. (\ref{aa}) is satisfied if, and only if, 
\begin{equation}
\delta P^{T} \cdot F^{T} \cdot F + F^{T} \cdot F \cdot \delta P + \delta X^{1} \dfrac{\partial f}{\partial X^{1}} \left( F^{T} \cdot F - I \right) = 0.
\end{equation}
Equivalently,
\begin{equation}\label{cc}
\left( \delta P^{T}+ \delta X^{1} \dfrac{\partial f}{\partial X^{1}}I \right) \cdot F^{T} \cdot F + F^{T} \cdot F \cdot \delta P  = \delta X^{1} \dfrac{\partial f}{\partial X^{1}}I.
\end{equation}
\noindent
The function in the left side of the equation is homogeneous of degree $2$ respect to the matrix coordinate $F$ but the function in the right side does not depend on $F$. So, Eq. (\ref{cc}) can be satisfied if, and only if,
\begin{equation}\label{dd}
\delta X^{1} \dfrac{\partial f}{\partial X^{1}}=0.
\end{equation}
Notice that, the map $f$ is strictly monotonic (and, hence, a submersion) at the open given by the condition $X^{1} >0$. Then, for any point $X$ in this open subset we have that
$$T_{X} f^{-1}\left( f \left( X^{1}\right)\right) = Ker\left( T_{X}f \right),$$
i.e., the tangent space of the level set $f^{-1}\left( f \left( X^{1}\right)\right) $, which is the plane $Y^{1}= X^{1}$, consists of vectors $V = \left( V^{1},V^{2},V^{3}\right)$ such that
$$V^{1} \left. \dfrac{\partial f}{\partial X^{1}} \right |_{X}=0.$$
In this way, a vector field $\Theta$ satisfies Eq. (\ref{aa}) if, and only if, $\delta P$ is skew-symmetric and the proyection $T \alpha \circ \Theta \circ \epsilon$ is tangent to the vertical planes $Y^{1} = C$. Therefore, for each point $X = \left( X^{1},X^{2},X^{3}\right)$ with $X^{1}>0$, the uniform leaf is given by the plane $Y^{1} = X^{1}$.\\
As a consequence, it is not hard to realize that the uniform leaf at the points satisfying $X^{1} = 0$ is, again, the plane $Y^{1}= 0$.\\
So, we conclude that a point $X = \left( X^{1},X^{2},X^{3}\right) \in \mathcal{B}_{0}$ is uniform of grade $3$ if $X^{1} <0$ and it is uniform of grade $2$ in another case.\\
Finally, the material body $\mathcal{B}_{0}$ is homogeneous. In fact, let us consider the canonical (global) coordinates $\left( X^{i} \right)$ of $\mathbb{R}^{3}$ restricted to $\mathcal{B}_{0}$. Then,
$$\dfrac{\partial W}{\partial X^{2}}  =  \dfrac{\partial W}{\partial X^{3}} = 0,$$
i.e., by using Proposition \ref{30}, $\mathcal{B}_{0}$ is homogeneous and the coordinates $\left( X^{i} \right)$ are homogeneous coordinates.

\subsection*{Example 2}\label{33}

We will consider a perturbation of the model introduced by Coleman \cite{COLE} and Wang \cite{CCWANTHIRD} called \textit{simple liquid crystal}. These kind of materials could be called \textit{laminated simple liquid crystals}.\\

In this case we will consider a simple body $\mathcal{B}$ together a reference configuration $\psi_{1}$ from $\mathcal{B}$ the open ball $\mathcal{B}_{1} = B_{r}\left( 0 \right)$ in $\mathbb{R}^{3}$ of radius $r$ and center $0 \in \mathbb{R}^{3}$. Furthermore, $\psi_{1}$ induces on $\mathcal{B}_{1}$ a mechanical response $\mathcal{W}$ determined by the following objects:
\begin{itemize}
\item[\textbf{(i)}]$\ $  A fixed vector field $e$ on $\mathcal{B}_{1}$ such that $e \left( X \right) \neq 0$ for all $X \in \mathcal{B}_{1}$. \\

\item[\textbf{(ii)}]$\ \ $  Two differentiable maps $r,J: \Pi^{1}\left( \mathcal{B}_{1} , \mathcal{B}_{1} \right) \rightarrow \mathbb{R}$ in the following way
\begin{itemize}
\item $r \left( j^{1}_{X,Y} \phi \right) = g \left( Y \right) \left( T_{X} \phi \left(  e \left( X  \right)\right),  T_{X} \phi \left( e \left( X  \right)\right) \right) + \parallel X \parallel^{2}$
\item $ J \left( j^{1}_{X,Y} \phi \right) = det \left( F \right)$
\end{itemize}
\noindent
where $F$ is the Jacobian matrix of $ \phi$ with respect to the canonical basis of $\mathbb{R}^{3}$ at $X$, $g$ is a Riemannian metric on $\mathcal{B}_{1}$ and $\parallel \cdot \parallel$ the euclidean norm of $\mathbb{R}^{3}$.\\

\item[\textbf{(iii)}]$\ \ $  A differentiable map $\widehat{W}: \mathbb{R}^{2} \rightarrow V$, with $V$ a finite-dimensional $\mathbb{R}-$vector space.
\end{itemize}
Thus, these three object induce a structure of simple body by considering the mechanical response $ \mathcal{W} : \Pi^{1} \left( \mathcal{B}_{1} , \mathcal{B}_{1} \right) \rightarrow V$ as the composition  
$$\mathcal{W} = \widehat{W} \circ  \left( r  , J \right).$$
Let us now fix the canonical (global) coordinates $\left(X^{i}\right)$ of $\mathbb{R}^{3}$. Then, these coordinates induce a (canonic) isomorphism $T \mathcal{B}_{1} \cong  \mathcal{B}_{1} \times \mathbb{R}^{3}$. By using this isomorphism any vector $V_{X} \in T_{X}\mathcal{B}_{1}$ can be equivalently expressed as $ \left( X , V^{i} \right)$ in $\mathcal{B}_{1} \times \mathbb{R}^{3}$. For the same reason, $r$ can be written as follows:
$$r \left( j^{1}_{X,Y} \phi \right) = g \left( Y \right) \left( F^{j}_{l} e^{l} \left( X  \right),  F^{j}_{l}e^{l} \left(X\right) \right) +  \parallel X \parallel^{2},$$
where $e \left( X \right) = \left( X, e^{i} \left( X \right) \right) \in \mathcal{B}_{1} \times \mathbb{R}^{3}$. Both expressions will be used with the same notation as long as there is no confusion.\\

Now, we want to study the conditions characterizing the material distribution $A \Omega^{T} \left( \mathcal{B}_{1} \right)$. In particular, we should study the admissible left-invariant vector fields $\Theta$ on $\Pi^{1} \left( \mathcal{B}_{1} , \mathcal{B}_{1} \right)$, i.e.,
\begin{equation}\label{a}
\Theta \left( \mathcal{W}\right) = 0.
\end{equation}
Notice that, for each $U= \left(U^{j}_{i}\right) \in gl \left( 3, \mathbb{R} \right)$ and $v=\left(v^{i}\right) \in \mathbb{R}^{3}$ we have that,
\begin{eqnarray*}
&\textbf{(i)}&\ \ \ \left. \dfrac{\partial r}{\partial X} \left( v \right) \right | _{j_{X,Y}^{1}\phi} =  2 \  g \left( Y\right)\left(  F^{j}_{l} e^{l} \left( X  \right) , F^{j}_{l} \left. \dfrac{\partial e^{l}}{\partial X^{m}} \right | _{X} v^{m}  \right) +\  2 \  v^{l}X^{l}.\\
&\textbf{(ii)}&\ \ \ \left. \dfrac{\partial r}{\partial F} \left(  U  \right) \right | _{j_{X,Y}^{1}\phi}= 2   \ g \left( Y\right)\left(F^{j}_{l} e^{l} \left( X  \right) ,U^{j}_{l} e^{l} \left( X  \right) \right).\\
&\textbf{(iii)}&\ \ \ \left. \dfrac{\partial J}{\partial F} \left(  U  \right) \right | _{j_{X,Y}^{1}\phi}= det \left( F \right) Tr \left( F^{-1} \cdot U \right).
\end{eqnarray*}
We are denoting the coordinate $X^{k}\left( X \right)$ by $X^{k}$.\\
Let $\left( X^{i} , Y^{j} , F^{j}_{i} \right)$ be the induced local coordinates of the canonical coordinates $\left( X^{i} \right)$ of $\mathbb{R}^{3}$ in $\Pi^{1} \left( \mathcal{B}_{1} , \mathcal{B}_{1} \right)$. Then, $\Theta$ can be expressed as follows,
$$ \Theta \left( X^{i} , Y^{j} , F^{j}_{i} \right) = \left( \left( X^{i} , Y^{j} , F^{j}_{i} \right)  , \delta X^{i} ,0,  F^{j}_{l} \delta P^{l}_{i} \right).$$
Hence, $\Theta$ is an admissible vector field if, and only if,

\begin{eqnarray*}
 0 &=& 2 \left. \dfrac{\partial \widehat{W}}{\partial r} \right | _{j_{X,Y}^{1}\phi} g \left( Y\right)\left( F^{j}_{l}   e^{l} \left( X  \right) ,F^{j}_{l}   \left. \dfrac{\partial e^{l}}{\partial X^{m}} \right | _{X} \delta X^{m} \left(X \right) \right) \\
&& +\  2 \ \left. \dfrac{\partial \widehat{W}}{\partial r} \right | _{j_{X,Y}^{1}\phi}  \delta X^{l} \left(X\right)  X^{l} \\
&&+\  2 \ \left. \dfrac{\partial \widehat{W}}{\partial r}\right | _{j_{X,Y}^{1}\phi}   g \left( Y\right)\left( F^{j}_{l}   e^{l} \left( X  \right) ,F^{j}_{l} \delta P^{l}_{m}\left(X \right) e^{m} \left( X  \right) \right)\\
&& +\   det \left(F \right) \left. \dfrac{\partial \widehat{W}}{\partial J} \right | _{j_{X,Y}^{1}\phi} Tr \left( \delta P^{j}_{i}\left(X\right)\right),
\end{eqnarray*}
for all $j_{X,Y}^{1} \phi \in \Pi^{1} \left( \mathcal{B}_{1} , \mathcal{B}_{1} \right)$. So, a sufficient but not necessary condition would be that for each $1-$jet of local diffeomorphisms $j_{X,Y}^{1} \phi $ on $ \mathcal{B}_{1}$ it satisfies that

\begin{eqnarray*}
&\textbf{(1)}&\ \ \ Tr \left( \delta P^{j}_{i}\left(X\right)\right) = 0.\\
&\textbf{(2)} &\ \ \ \mbox{\small $g \left( Y\right)\left( F^{j}_{l}   e^{l} \left( X  \right)  , F^{j}_{l}    \left( \left. \dfrac{\partial e^{l}}{\partial X^{m}} \right | _{X} \delta X^{m} \left(X\right)+ \delta P^{l}_{m}  \left(X\right) e^{m} \left( X \right) \right)\right)$}\\ &\ \ \  &\ \ \ \mbox{\small $= - \delta X^{l} \left(X\right) X^{l}$},
\end{eqnarray*}

In order to turns this conditions into necessary conditions we will assume that $\widehat{W}$ is an immersion and, hence, \textbf{(1)} and \textbf{(2)} are equivalent to Eq. (\ref{a}).\\
In this way, $\mathcal{B}_{1}$ is smoothly uniform if, and only if, for each vector $V_{X}$ at $X$ there exists a family of local functions $\{ \delta X^{i} , \delta P^{i}_{j} \}$ at $X$ satisfying \textbf{(1)}, \textbf{(2)} and
$$ \delta X^{i} \left( X\right) = V^{i}, \ \forall i$$
where $V_{X} = \left( X , V^{i}\right) \in \mathcal{B}_{1} \times \mathbb{R}^{3}$.\\
Let us focus on the second condition: Suppose that $\left\langle V^{i} , X \right\rangle  \neq 0$. Then, fixing the spatial point $X \in \mathcal{B}_{1}$ the map depending on the matrix coordinates $F^{j}_{i}$,
\begin{equation}\label{b}
g \left( Y\right)\left( F^{j}_{l}   e^{l} \left( X  \right)  , F^{j}_{l}  \left(   \left. \dfrac{\partial e^{l}}{\partial X^{m}} \right | _{X} V^{m} + \delta P^{l}_{m}  \left(X\right) e^{m} \left( X \right) \right)\right),
\end{equation}
is equal to $- \delta X^{l} \left(X\right) X^{l}$ which does not depend on the matrix coordinates $F^{j}_{i}$ and it is not zero. However, the map (\ref{b}) depends bilinearly on $F^{j}_{i}$. So, (\ref{b}) cannot be constant (respect to $F^{j}_{i}$) and different to zero at the same time. Therefore, we could conclude that $\mathcal{B}_{1}$ is not smoothly uniform.\\
This fact opens the possibility of studying the graduated uniformity of these materials. Notice that, as we have proved, any admissible vector field $\Theta$ satisfies that
\begin{equation}\label{c}
\delta X^{l} \left( X \right) X^{l} = 0,
\end{equation}
where $\Theta \left( X^{i}, Y^{j},F^{j}_{i}\right) =  \left( \left( X^{i} , Y^{j} , F^{j}_{i} \right)  , \delta X^{i} ,0,  F^{j}_{l} \delta P^{l}_{i} \right)$ respect to the coordinates $\left( X^{i} , Y^{j} , F^{j}_{i} \right)$ on $\Pi^{1} \left( \mathcal{B}_{1} , \mathcal{B}_{1} \right)$.\\
Let $X \in \mathcal{B}_{1}$ be a point of the body different to $0$. Then, the map given by $ \parallel \cdot \parallel^{2}$ restricted to $\mathcal{B}_{1}$ has full rank at $X$. In fact, the level set of $ \parallel \cdot \parallel^{2}$ at $ \parallel X \parallel^{2}$ is given by the sphere $S\left( \parallel  X  \parallel \right)$ of radius $\parallel X \parallel$ and centre $0$ and it satisfies that
$$T_{X} S\left( \parallel  X  \parallel \right) = Ker \left( T_{X} \parallel \cdot \parallel^{2}\right).$$
So, the tangent space of the sphere $S\left( \parallel  X  \parallel \right)$ at $X$ consists of the vectors $V_{X} = \left( X , V^{i} \right) \in \mathcal{B}_{1} \times \mathbb{R}^{3}$ satisfying
\begin{equation}\label{i}
V^{l}X^{l} = 0
\end{equation}
Then, any vector $V_{X}$ satisfying Eq. (\ref{i}) can be expanded by a (local) vector field $\theta^{S}$ on $S\left( \parallel  X  \parallel \right)$ such that
$$ \theta^{S} \left( X \right)= V_{X}.$$
It now an easy exercise to prove that $\theta^{S}$ can be extended to a vector field $\theta$ on an open neighbourhood $U$ of $\mathcal{B}_{1}$ which is tanget to all the spheres intersecting $U$. Then, expressing $\theta$ in the canonical coordinates $\left(X^{i}\right)$ as follows
$$\theta \left( X^{i}\right) = \left( X^{i} , \delta X^{i} \right),$$
the functions $\delta X^{i}$ satisfy Eq. (\ref{c}). Therefore, by using the non-degenerance of the Riemannian metric $g$, it is enough to realize that there exist infinite families of local maps $\delta P^{j}_{i}$ at $X$ from the body to $\mathbb{R}$ satisfying that
\begin{eqnarray*}\label{d}
&\textbf{(1)''}&\ \ \   \delta P^{i}_{i}= 0.\\
&\textbf{(2)''}&\ \  \dfrac{\partial e^{j}}{\partial X^{l}}   \delta X^{l} = -\delta P^{j}_{l}  e^{l} , \ \forall j,
\end{eqnarray*}
Therefore, the local vector fields given by,
$$\Theta \left( X^{i}, Y^{j},F^{j}_{i}\right) =  \left( \left( X^{i} , Y^{j} , F^{j}_{i} \right)  , \delta X^{i} ,0,  F^{j}_{l} \delta P^{l}_{i} \right),$$
satisfy Eq. (\ref{a}) and $\delta X^{i} \left( X\right) = V_{X}$. Then, we have already proved that the grade of uniformity of any point $X$ at $\mathcal{B}_{1}$ different to $0$ is $2$ and the smoothly uniform submanifolds are given by the spheres $S \left( \parallel  X  \parallel \right)$. Then, obviously, the grade of uniformity of $0$ is $0$ and the smoothly uniform submanifold at $0$ is $\{0\}$. Therefore, ignoring the zero, $\mathcal{B}_{1}$ is a ``laminated" body covered by smoothly uniform submanifolds of dimension $2$ with a kind of structure similar to \textit{liquid crystals}.\\
Let us now test the (local) \textit{homogeneity} of $\mathcal{B}$. In this sense, by using again Proposition \ref{30}, we should study the existence of a system of (local) coordinates $\left( x^{i} \right)$ at each $X \in \mathcal{B}_{1}$ such that,
\begin{equation}\label{e}
 \dfrac{\partial \mathcal{W}}{\partial x^{l}}= 0,
\end{equation}
for all $l\leq 2$ if $X \neq 0$ and $l=0$ if $X = 0$.\\
Let $\left(x^{i}\right)$ be a system of local coordinates of $\mathcal{B}_{1}$. Using the chain rule we have that,
$$\dfrac{\partial \mathcal{W}}{\partial x^{i}} = \dfrac{\partial \widehat{W}}{\partial r} \dfrac{\partial r}{\partial x^{i}} + \dfrac{\partial \widehat{W}}{\partial J} \dfrac{\partial J}{\partial x^{i}}.$$

Therefore, the immersion property of $\widehat{W}$ implies that $\left( x^{i} \right)$ are homogeneous coordinates if, and only if,
\begin{eqnarray*}
&\textbf{(1)'''}&\ \ \ \dfrac{\partial r}{\partial x^{l}} = 0.\\
&\textbf{(2)'''}&\ \ \ \dfrac{\partial J}{\partial x^{l}} = 0. 
\end{eqnarray*}
for all $l\leq 2$ if $X \neq 0$ and $l=0$ if $X = 0$. Hence, the study of homogeneity depends only on the properties of $r$ and $J$.\\
For each $j_{X,Y}^{1} \phi \in \Pi^{1} \left( \mathcal{B}_{1} , \mathcal{B}_{1} \right)$
\begin{eqnarray*}
r \left( j_{X,Y}^{1} \phi \right) & = & g \left( Y\right)\left( T_{X}\phi \left( e \left( X \right)\right) ,  T_{X}\phi \left( e \left( X \right)\right) \right) + \parallel X \parallel^{2}\\
& = &  \mbox{\footnotesize $g \left( Y\right)\left( T_{X}\phi \left( e^{i} \left( X \right)\left. \dfrac{\partial }{\partial x^{i}}\right | _{X}\right) ,  T_{X}\phi \left( e^{j} \left( X \right)\left. \dfrac{\partial }{\partial x^{j}}\right | _{X}\right) \right) + \parallel X \parallel^{2}$}\\
& = & e^{i} \left( X \right)e^{j} \left( X \right) \left. \dfrac{\partial \phi^{k}}{\partial x^{i}}\right | _{X} \left. \dfrac{\partial \phi^{l} }{\partial x^{j}} \right | _{X} g_{kl} \left( Y \right) + \parallel X \parallel^{2}
\end{eqnarray*}
where, in this case, $e^{j}$ are the coordinates of $e$ respect to $\left(x^{i} \right)$. So, considering the induced coordinates $\left( x^{i} , y^{j} , y^{j}_{i}\right)$ of $\left( x^{i} \right)$ on $\Pi^{1} \left( \mathcal{B}_{1} , \mathcal{B}_{1} \right)$ we have that for all $\left( \tilde{X} , \tilde{Y} , \tilde{F} \right)$,
\begin{eqnarray*}
r \circ \left( x^{i} , y^{j} , y^{j}_{i}\right)^{-1} \left( \tilde{X} , \tilde{Y} , \tilde{F} \right) & = & e^{i} \left( X \right)e^{j} \left( X \right)\tilde{F}^{k}_{i}\tilde{F}^{l}_{j} g_{kl} \left( Y \right) + \left( X^{i}\right)^{2},\\
\end{eqnarray*} 
where $X = \left( x^{i}\right)^{-1}\left( \tilde{X} \right)$ and $Y = \left( x^{i}\right)^{-1}\left( \tilde{Y} \right)$. In this way, 

\begin{eqnarray*}
\left. \dfrac{\partial r}{\partial x^{k}} \right | _{j_{X,Y}^{1}\phi} & = & 2 \left. \dfrac{\partial e^{i}}{\partial x^{k} } \right | _{X} e^{j} \left( X \right)\tilde{F}^{k}_{i}\tilde{F}^{l}_{j} g_{kl} \left( Y \right) + 2 X^{l} \left. \dfrac{\partial X^{l}}{\partial x^{k} } \right | _{X}.\\
\end{eqnarray*}
So, let us study the equation,
$$ \dfrac{\partial r}{\partial x^{k}} = 0.$$
Again, the dependence of the matrix variable on the left side of the equations take us to the necessary equation,
$$
X^{l} \left. \dfrac{\partial X^{l}}{\partial x^{k} } \right | _{X} =0
$$
Hence, by using the non-degeneracy of $g$ we have that $ \dfrac{\partial r}{\partial x^{k}} = 0$ if, and only if,
\begin{itemize}
\item[\textbf{(i)}]
\begin{equation}\label{k}
X^{l}  \dfrac{\partial X^{l}}{\partial x^{k} } =0
\end{equation}

\item[\textbf{(ii)}]
\begin{equation}\label{f}
\dfrac{\partial e^{i}}{\partial x^{k}} = 0, \ \forall i.
\end{equation}
\end{itemize}
Thus, $\textbf{(1)'''}$ is satisfied if, and only if,
\begin{equation}\label{g}
\dfrac{\partial e^{i}}{\partial x^{l}} = 0, \ \ \ \ \ \ \ \ \ \ \ \ \ \ \ \ \ \ \ \ \ X^{r}  \dfrac{\partial X^{r}}{\partial x^{l} } =0, \ \ \ \  \forall i, \ l\leq 2.
\end{equation}
These two equations can be translated as that the functions $e^{i}$ are constants on the spheres and the partial derivatives $\dfrac{\partial}{\partial x^{i} }$ are tangent to the spheres.\\
Next, let us study condition $\textbf{(2)'''}$. Notice that,

\begin{scriptsize}
\begin{eqnarray*}
\left[ J \circ \left( x^{i} , y^{j} , y^{j}_{i}\right)^{-1} \right] \left( \tilde{X} , \tilde{Y} , \tilde{F}  \right) &=& J  \left( j_{ \left( \left( x^{i}\right)^{-1}\left( \tilde{X} \right) , \left( y^{j}\right)^{-1}\left( \tilde{Y} \right)\right)}^{1} \left( \left( y^{j}\right)^{-1} \circ \tilde{\phi} \circ \left( x^{i}\right) \right)\right)\\
&=& J \left( \nabla_{\tilde{Y}}\left( y^{j}\right)^{-1}\right) J \left( \tilde{ F} \right) J \left( \nabla_{\left( x^{i}\right)^{-1}\left( \tilde{X} \right)} \left(x^{i}\right)\right),
\end{eqnarray*}
\end{scriptsize}
\noindent{where $\tilde{F}= \nabla_{\tilde{X}} \tilde{\phi}$ and $\left( \tilde{X} , \tilde{Y} , \tilde{F}  \right)$  is in the codomain of $\left( x^{i} , y^{j} , y^{j}_{i}\right)$. Then, $ \dfrac{\partial J}{\partial x^{i}} = 0$ if, and only if,}
\begin{equation}\label{j}
\dfrac{\partial}{\partial X^{i}} \left( J \left( \nabla_{\left( x^{i}\right)^{-1}\left( \tilde{X} \right)} \left(x^{i}\right)\right)\right) = 0.
\end{equation}
So, denoting $\left( x^{i}\right)^{-1}\left( \tilde{X} \right)$ by $X$, Eq. (\ref{j}) is equivalent to
$$  \left. \dfrac{\partial^{2}  x^{k}}{\partial X^{i}\partial X^{m}}\right | _{X} = 0, \ \forall k,m.$$
Therefore, $\textbf{(2)'''}$ is can be expressed as follows,
\begin{equation}\label{h}
 \dfrac{\partial^{2}  x^{i}}{\partial X^{k}\partial X^{m}} = 0, \ \forall i,m, \ \forall k \leq 2.
\end{equation}
We conclude with this that $\mathcal{B}_{1}$ (locally) homogeneous if, and only if, there exists a local system of coordinates $\left( x^{i} \right)$ which satisfies that the functions $e^{i} = e \left( x^{i}\right)$ are constants on the spheres, the partial derivatives $\dfrac{\partial}{\partial x^{i} }$ are tangent to the spheres and it satisfies Eq. (\ref{h}).\\
Therefore, in general, $\mathcal{B}_{1}$ is not (locally) homogeneous. In fact, let us consider the following vector field
$$e =  X^{k} \dfrac{\partial}{\partial X^{k}} + r\dfrac{\partial}{\partial X^{1}}.$$
The factor $r\dfrac{\partial}{\partial X^{1}}$ is added to get that the vector field $e$ does not vanish.\\
Assume that there exists a local system of homogeneous coordinates $\left( x^{i} \right)$ on $\mathcal{B}_{1}$. Notice that, 
$$ e = \left( X^{k} \right) \dfrac{\partial x^{i}}{\partial X^{k}}\dfrac{\partial}{\partial x^{i}} + r\dfrac{\partial x^{i}}{\partial X^{1}}\dfrac{\partial}{\partial x^{i}},$$
i.e., the coordinates $e^{i}$ respect to the coordinates $\left( x^{i}\right)$ are given by $ X^{k} \dfrac{\partial x^{i}}{\partial X^{k}} + r\dfrac{\partial x^{i}}{\partial X^{1}}$. Then, it should satisfy that for each $l \leq 2$
\begin{eqnarray*}
\dfrac{\partial e^{i}}{\partial x^{l}} &=& \dfrac{\partial}{\partial x^{l}}\left( X^{k}  \dfrac{\partial x^{i}}{\partial X^{k}}\right) + r\dfrac{\partial^{2} x^{i}}{\partial x^{l} \partial X^{1}}\\
&=& \dfrac{\partial X^{k} }{\partial x^{l}} \dfrac{\partial x^{i}}{\partial X^{k}} +   X^{k}\dfrac{\partial^{2} x^{i}}{\partial x^{l}\partial X^{k}} + r\dfrac{\partial^{2} x^{i}}{\partial x^{l} \partial X^{1}}\\
&=& \delta^{i}_{l} +   X^{k}\dfrac{\partial^{2} x^{i}}{\partial x^{l}\partial X^{k}} + r\dfrac{\partial^{2} x^{i}}{\partial x^{l} \partial X^{1}}.
\end{eqnarray*}
Notice that,
\begin{eqnarray*}
\dfrac{\partial^{2} x^{i}}{\partial x^{l} \partial X^{1}} &=& \dfrac{\partial}{\partial X^{l}} \left(  \dfrac{\partial x^{i}}{ \partial X^{1}} \circ \left(x^{i}\right)^{-1} \right)\\
&=& \dfrac{\partial^{2} x^{i}}{\partial X^{k} \partial X^{1}} \dfrac{\partial X^{k}}{\partial x^{l}} =0.
\end{eqnarray*}
This is a consequence of Eq. (\ref{h}). So,
$$\dfrac{\partial e^{i}}{\partial x^{l}} = 0,$$
if, and only if, 
\begin{equation}\label{l}
\delta^{i}_{l} = - X^{k}\dfrac{\partial^{2} x^{i}}{\partial x^{l}\partial X^{k}}.
\end{equation}
Notice that, Eq. (\ref{l}) implies that for $i \neq l$,
$$X^{k}\dfrac{\partial^{2} x^{i}}{\partial x^{l}\partial X^{k}} = 0.$$
Thus, considering $X= \left( 0,0, c \right)$ with $c \neq 0$ we have that
\begin{eqnarray*}\label{m}
\left.\dfrac{\partial^{2} x^{i}}{\partial x^{l}\partial X^{3}} \right | _{X}&=& \left. \dfrac{\partial^{2} x^{i}}{\partial X^{k} \partial X^{3}} \right | _{X}  \left. \dfrac{\partial X^{k}}{\partial x^{l}} \right | _{X}\\
&=& \left. \dfrac{\partial^{2} x^{i}}{\partial X^{3} \partial X^{3}} \right | _{X}  \left. \dfrac{\partial X^{3}}{\partial x^{l}} \right | _{X} = 0.
\end{eqnarray*}
Then $ \left. \dfrac{\partial^{2} x^{i}}{\partial X^{3} \partial X^{3}} \right | _{X} =0$ or $   \left. \dfrac{\partial X^{3}}{\partial x^{l}} \right | _{X} =0$. Observe that, by Eq. (\ref{l}), for each $l \leq 2$
$$1 = \left. \dfrac{\partial^{2} x^{l}}{\partial X^{3} \partial X^{3}} \right | _{X}  \left. \dfrac{\partial X^{3}}{\partial x^{l}} \right | _{X}.$$
So, the expression $   \left. \dfrac{\partial X^{3}}{\partial x^{l}} \right | _{X}$ cannot be zero. For the same reason, for $i \leq 2$, $ \left. \dfrac{\partial^{2} x^{i}}{\partial X^{3} \partial X^{3}} \right | _{X} $ is different to $0$. Therefore, Eq. (\ref{m}) cannot be satisfied and the laminated simple liquid crystal $\mathcal{B}_{1}$ induced by this vector field $e$ is not homogeneous.\\

\bibliographystyle{plain}
\bibliography{Library}

\end{document}